\documentclass[fleqn,usenatbib]{mnras}

\usepackage{newtxtext,newtxmath}

\usepackage[T1]{fontenc}
\usepackage{ae,aecompl}
\usepackage[usenames,dvipsnames,svgnames,hyperref]{xcolor}
\usepackage{soul,ulem}

\usepackage{graphicx}	
\usepackage{amsmath}	
\usepackage{subfigure}
\usepackage{multirow}
\usepackage{enumitem}

\usepackage{tikz}
\usetikzlibrary{arrows.meta}
\tikzset{>={Latex[width = 1mm,length = 2mm]},
    base/.style = {rectangle, draw = blue, thick, minimum width = 8cm, minimum height = 1cm, text centered, font = \sffamily}, 
    repeat1/.style = {base, rounded corners, draw = red, minimum width = 7cm}, 
    repeat2/.style = {repeat1, draw = orange, minimum width = 6cm}
}

\def\red{\textcolor{red}}

\def\code#1{{\tt \textsc{#1}}}
\def\clustar{\code{CluSTAR-ND}}
\def\hoptics{\code{Halo-OPTICS}}
\def\optics{\code{OPTICS}}

\usepackage{stfloats}

\title[Hierarchical Clusters in Haloes]{The Hierarchical Structure of Galactic Haloes: 
\\
Generalised N-Dimensional Clustering with \textsc{CluSTAR-ND}}
\author[Oliver et al.]{
William H. Oliver$^{1}$\thanks{E-mail: william.oliver@sydney.edu.au},
Pascal J. Elahi$^{2}$, and
Geraint F. Lewis$^{1}$
\\
$^{1}$Sydney Institute for Astronomy, School of Physics A28, The University of Sydney, NSW, 2006, Australia\\
$^{2}$Pawsey Supercomputing Research Centre, 1 Bryce Avenue, Kensington, WA, 6151, Australia
}

\date{Accepted XXX. Received YYY; in original form ZZZ}

\pubyear{2022}

\begin{document}
\label{firstpage}
\pagerange{\pageref{firstpage}--\pageref{lastpage}}
\maketitle

\begin{abstract}
We present \clustar, a fast hierarchical galaxy/(sub)halo finder that produces {\bf Clu}stering {\bf S}tructure via {\bf T}ransformative {\bf A}ggregation and {\bf R}ejection in {\bf N}-{\bf D}imensions. It is designed to improve upon \hoptics\ -- an algorithm that automatically detects and extracts significant astrophysical clusters from the 3D spatial positions of simulation particles -- by decreasing run-times, possessing the capability for metric adaptivity, and being readily applicable to data with any number of features. We directly compare these algorithms and find that not only does \clustar\ produce a similarly robust clustering structure, it does so in a run-time that is at least $3$ orders of magnitude faster. In optimising \clustar's clustering performance, we have also carefully calibrated $4$ of the $7$ \clustar\ parameters which -- unless specified by the user -- will be automatically and optimally chosen based on the input data. We conclude that \clustar\ is a robust astrophysical clustering algorithm that can be leveraged to find stellar satellite groups on large synthetic or observational data sets.
\end{abstract}




\section{Introduction} \label{sec:introduction}
The process of identifying clusters -- often referred to as clustering\footnote{Note that we use this term throughout this paper to mean both the general process of finding clusters and the resultant set of classifications from this process -- depending on the context. As such, our use of this term is much broader than its typical use within astro- and cosmo-related fields, e.g. referring specifically to large-scale-structure.} -- from a data set has been an ongoing data mining problem within the field of machine learning. The function of any algorithm tasked with undertaking this problem is to determine a set of statistically coherent groups within the intrinsic feature space of the data. Many such algorithms exist for this purpose, however due to the innate subjectivity of the definition of a {\it cluster} it is not always obvious as to which of these may be useful for clustering of a given type.

The groupings found by a clustering algorithm can be categorised into one or more of the typical models, such as; centroid-based \citep[e.g. {\code{K-Means}};][]{MacQueen1967, Lloyd1982}, distribution-based \citep[e.g. {\code{EM}};][]{Dempster1977}, density-based \citep[e.g. {\code{DBSCAN}};][]{Ester1996}, and others. The way in which a clustering algorithm partitions the data is also of importance and provides a similar categorisation. These algorithms may return a flat or hierarchical clustering such that clusters are mutually exclusive or can be proper sub/supersets of one another -- this is one of the differences between \code{DBSCAN} and \code{HDBSCAN} \citep{Campello2015} for example. In addition to these, an algorithm may also return an overlapping set of clusters. A clustering may place all points within clusters (e.g. \code{K-Means}) or more commonly it may leave some points out of clusters classifying them as noise or outliers to the clusters that have been predicted. Moreover, these algorithms may return a hard or soft -- also referred to as fuzzy -- clustering. This distinguishment clarifies whether the nature of a point being within a cluster is binary-based or probability-based \citep[e.g {\code{Fuzzy C-Means}};][]{Dunn1973}. Since each of these separate class systems are codependent -- in that an algorithm may be classified by multiple at a time -- it suffices to say that choosing an appropriate clustering algorithm for the problem at hand is non-trivial.

In cosmology, a typical description of a galaxy and its halo is any spatial overdensity that is denser than the critical -- or mean -- density of the universe by some factor $\Delta$. Two common ways to identify such overdensities from both cosmological simulations and observed data sets are via the Spherical-Overdensity method \citep[{\code{SO}};][]{Press1974} and the Friends-Of-Friends algorithm \citep[{\code{FOF}};][]{Davis1985}. By themselves, these algorithms perform a density-based flat and hard clustering with noise which is well suited to the intended definition of galactic haloes -- and as an additional constraint, the \code{SO} method can also be considered part of the distribution-based family due to ensuring that clusters adhere to a particular volumetric shape.

It is a primary prediction of the $\Lambda$CDM cosmological model that galaxies are formed hierarchically via continual accretion and merger events \citep{White1978, Kauffmann1993, Ghigna1998}. Depending upon the conditions of the satellite at the time of the merger as well as the ongoing conditions of the host halo \citep{Bullock2005, Johnston2008}, the particles within these infalling groups may become mixed in with, and indistinguishable from, the surrounding halo\footnote{The phase-space volume of infalling groups is conserved in a fully collisionless Newtonian gravity simulation.}. This prediction is mostly confirmed in nature barring a few exceptions such as the missing satellite problem \citep{Klypin1999, Moore1999, Reed2005, Tollerud2008, Springel2008, Ishiyama2013} and the core-cusp problem \citep{Flores1994, Moore1994, VanDenBosch2000}. As such, an appropriate flavour of clustering algorithm for application to within galactic haloes is that of the density-based hierarchical type with noise. Many such algorithms exist for specifically this purpose, including but not limited to; \code{SUBFIND} \citep{Springel2001}, \code{AHF} \citep{Knollmann2009}, \code{ROCKSTAR} \citep{Behroozi2012}, \code{VELOCIraptor} \citep{Elahi2019}, and more. Comparisons of these and others tend to show the outputs of halo-finders to be similar \citep{Knebe2011, Knebe2013}.

The above halo-finders mostly rely on the \code{SO} method and/or the \code{FOF} algorithm which both restrict the clustering to certain types of clusters. The varied shapes and density profiles of subhaloes often mean that these base algorithms need to be applied iteratively in order to find a range of subhalo types. In doing so, the run-times become larger although this does also allow some of these halo-finders to return hierarchical clusterings and even utilise adaptive metrics. Despite these measures of fine-tuning, \code{SO}-based halo-finders struggle to find curved structures such as streams and \code{FOF}-based halo-finders can still fall victim to point-point noise and fail to detect sparsely populated structures. The galaxy/(sub)halo finder \hoptics\ \citep{Oliver2020} is based on \optics\ \citep{Ankerst1999}, a generalisation of \code{DBSCAN}, and is more robust to point-point noise. Coupled with an adaptive metric, \hoptics\ could be a powerful astrophysical clustering algorithm if not for the fact that the run-time of \optics\ can become unmanageable when operating in this way. Alternatives typically only use a set of nearest neighbours to perform locally adaptive metric computation. One such algorithm, the subhalo-finder \code{EnLink} \citep{Sharma2009}, uses an entropy-based locally adaptive metric in combination with a group finder that is also robust to point-point noise in order to find a range of subhalo types. \code{VELOCIraptor} and \code{HSF} \citep{Maciejewski2009} also can attempt to generate locally adaptive metrics to identify clusters. 

Having a fast clustering algorithm is not simply a matter of convenience. It is critical when clustering over extensive data sets with large feature spaces. It is also of particular importance when used to find a fuzzy clustering of an astrophysical data set. Unlike the output from the \code{Fuzzy C-Means} algorithm, which is a fuzzy clustering of a hard data set, the fuzzy clustering found in this context is itself from a fuzzy data set and needs to have propagated the uncertainties of a point's features into the probability of that point's membership within a given cluster as well as the probability of that cluster's existence. Solving this problem typically requires taking many samplings of the data and comparing each of the clusterings therefrom \citep[e.g.][]{Fuentes2017, Malhan2022}. In order to be able to perform such a task whilst also maintaining a high-calibre clustering power, it is becoming necessary to use an algorithm that is: density-based; hierarchical; robust to point-point noise; equipped with adaptive metric capabilities; applicable to data sets with any number of features; and demonstrates exceptionally modest run-times.

We present \clustar, a derivative of \hoptics\ that possesses the above qualities, and apply it to synthetic survey data produced by \code{Galaxia} \citep{Sharma2011} in order to formalise the process of clustering n-dimensional data sets of galactic haloes. We first provide a relevant outline of the \hoptics\ algorithm (Sec. \ref{sec:halooptics}). We then describe the \clustar\ algorithm in Sec. \ref{sec:clustar}. In doing so we give an overview of the concept and motivation behind the algorithm (Sec. \ref{subsec:concept}), describe the root-haloes produced (Sec. \ref{subsec:rootclusters}), and the means by which the substructure therein is found (Secs. \ref{subsec:transformation} -- \ref{subsec:rejection}). In Sec. \ref{sec:syntheticdata} we summarise the details of the synthetic galaxies from \code{Galaxia}. Following this we make a direct comparison between \clustar\ and the \hoptics\ algorithm in Sec. \ref{sec:comparison}. In Sec. \ref{sec:optimisation} we discuss the influence of the algorithm's parameters and optimise them by training them on the galaxies from \code{Galaxia}. We then discuss the contrast in the information content available from the clusterings produced from different clustering scenarios in Sec. \ref{sec:infocontent}. Finally, we present our conclusions and directions for future work in Sec. \ref{sec:conclusion}.

\section{An Overview of the \textsc{Halo-OPTICS} Algorithm} \label{sec:halooptics}
In order to understand \clustar\ it is first necessary to grasp the algorithm \hoptics\ \citep{Oliver2020}. \hoptics\ is a hierarchical galaxy/(sub)halo finder that can be used for the density-based determination of astrophysical clusters in a 3D spatial data set via a global distance metric, i.e. the Euclidean distance. It is an extension of the well-known hierarchical clustering algorithm, \optics\ \citep{Ankerst1999} -- which is itself an extension of \code{DBSCAN} \citep{Ester1996}. \code{DBSCAN} is non-hierarchical and will produce a flat clustering with noise of a given data set such that the points in each cluster are {\it densely connected} with a density greater than some threshold. All points with a density lower than this threshold are classified as noise and are not clustered. \optics\ not only connects all points in this manner, but also keeps track of how the points are connected together -- i.e. in which order and with what measure of local density. This adjustment allows \optics\ to build a {\it reachability plot} which contains information of the clustering structure.

Both the \optics\ and \code{DBSCAN} algorithms require $2$ parameters; $\epsilon$, a search radius, and $N_{\mathrm{min}}$ (often denoted as {\it MinPts}), the minimum number of points a cluster can have. The algorithms also need a distance metric to be defined over the feature space of the input data. Conditioned upon this metric, \optics\ uses the concepts of {\it core distance} and {\it reachability distance} to produce its output -- the reachability plot. The core distance of any given point is the distance between that point and it's ${N_{\mathrm{min}}}^{\mathrm{th}}$ nearest neighbour. It then follows that a {\it core-point} is any point whose core distance is less than or equal to $\epsilon$. The reachability distance of any point, $q$, with respect to another, $o$, is the maximum of the core distance of $o$ and the distance between $q$ and $o$.

Initially, \optics\ computes the core distances of each point and sets the reachability distances of all points to infinity. The algorithm then iteratively orders each point in the data set. In each iteration of this ordering process, the next-to-be-ordered point is chosen as the point with the smallest reachability distance from the set of all unordered points -- for this reason the first point is chosen at random. The chosen point is then removed from the list of unordered points and appended to the ordered list. If this point is a core-point, then all remaining unordered points within a radius of $\epsilon$ of it are found. The reachability distance of each of these unordered neighbours is then set to be the minimum of their currently assigned reachability distance and their reachability distance with respect to the recently ordered core-point. Following this step, the ordering process continues on to the next iteration -- repeating this cycle until all points are in the ordered list. It is in this manner that \optics\ creates an ordering of the data points, whilst concurrently seeking out regions of minimal reachability distance, i.e. highest density. After the ordering process, the reachability plot can then be constructed by plotting the final reachability distances as a function of the corresponding ordered indices, for all points. Within this plot, mutually clustered points appear as valleys since these points are both denser than their surrounds (smaller distances between points) and local to each other (ordered consecutively). More details on the \optics\ algorithm can be found in \citet{Ankerst1999} and \citet{Oliver2020}.

\hoptics\ takes the \optics\ algorithm and not only formalises the way in which the \optics\ parameters are chosen but also provides a robust cluster extraction technique that operates on the reachability plot in order to extract a tree-structured hierarchy of suitable astrophysical clusters\footnote{The tree-structure of this hierarchy allows for clusters to be referred to by the typical terminology, e.g. root-level cluster (largest cluster in the tree), parent cluster (superset of the child), child cluster (subset of the parent), and leaf cluster (has no child clusters of its own).}. In \hoptics, the \optics\ parameter $\epsilon$ is effectively converted to the more physical parameter $\Delta$, thereby classifying the root-level galaxy haloes using the factor by which they are denser than the critical (or mean) density of the Universe. \hoptics\ does this by using an approximate mapping between the \code{FOF} linking length $l_x$, $N_{\mathrm{min}}$, and $\epsilon$. The technique of extracting clusters is based on the designs of \citet{Sander2003}, \citet{Zhang2013}, and \citet{McConnachie2018}. The cluster extraction technique of \hoptics\ creates the hierarchy of clusters by finding the valleys within valleys that satisfy a range of conditions. These conditions assert that a cluster must; contain at least $N_{\mathrm{min}}$ points; have a median density that is a factor of $\rho_{\mathrm{threshold}}$ denser than the cluster's surrounds; not be a single child cluster of it's parent cluster; not share more than $f_{\mathrm{reject}}$ of it's points with it's parent; and it rejects local outliers that have local-outlier-factors (defined in Eq. \ref{eq:lof} and \citet{Breunig1999}) greater than or equal to $S_{\mathrm{outlier}}$ from each cluster. The \hoptics\ algorithm therefore transforms one \optics\ parameter and adds $3$ unitless parameters for which near optimal values are $\rho_{\mathrm{threshold}} = 2$, $f_{\mathrm{reject}} = 0.9$, and $S_{\mathrm{outlier}} = 2$. More details on the \hoptics\ algorithm can be found in \citet{Oliver2020}.

\begin{figure}
\begin{center}
\begin{tikzpicture}[node distance=1.5cm, every node/.style={fill=white, text centered, font=\sffamily}], align=center]
    \node[align = center] (input)[base]
    {Input the data and parameters.};
    
    \node[align = center] (fieldhaloes)[base, below of=input]
    {Find all field haloes using 3D \code{FOF} or treat\\input data as if it were a field halo.};
    
    \node[align = center] (searchforsubstructure)[base, below of=fieldhaloes]
    {For each field halo, search for substructure\\within it using the process in Fig. \ref{fig:searchsubstructure}.};
    
    \node[align = center] (output)[base, below of=searchforsubstructure]
    {Output the hierarchy of clusters.};
    
    \draw[->](input) -- (fieldhaloes);
    \draw[->](fieldhaloes) -- (searchforsubstructure);
    \draw[->](searchforsubstructure) -- (output);
\end{tikzpicture}
\end{center}
\vspace{-8pt}
\caption{An activity of the outer methods of the \clustar\ process. Before finding the substructure \clustar\ first decides on the root-level clusters - where the option is given to produce 3D \code{FOF} field haloes or to treat the input data as if it were a field halo.}
\label{fig:rootlevel}
\end{figure}
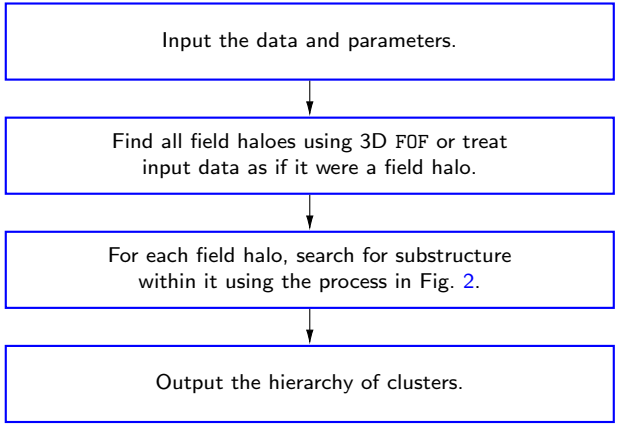

\section{CluSTAR-ND: A Hierarchical Galaxy/(Sub)Halo Finder} \label{sec:clustar}
\subsection{Concept and Motivation} \label{subsec:concept}
Originally, our intention was to create an optimised version of the \hoptics\ algorithm (refer to Sec. \ref{sec:halooptics} for details) that implemented a locally adaptive metric and could be applied to a data set with an arbitrary number of features with minimal additional input from the user. Ultimately, since \hoptics\ iteratively performs a radial search about each of the $n$ $d$-dimensional points in the data set, the cost of having it compute the distances via a locally adaptive metric is too great. For $m$ points within some radius of the query point, this algorithm would effectively need to have computed the inverse and determinant of $m$-$d \times d$ covariance matrices which makes the time complexity of the radial search and distance calculations $\mathcal{O}(m\mathrm{log}(n))$ and $\mathcal{O}(md^{3})$ respectively (each with different constant factors). In practice this increases the run-times dramatically such that even applying this to small galactic haloes becomes challenging.

We have designed \clustar\ with the intention of mimicking \hoptics\ along with the additional benefits of having; faster run-times, a variety of adaptive metric settings, and the capability of performing on a data set with any number of features. In addition to the input data, \clustar\ has $7$ parameters that it uses to construct clusters:

\begin{enumerate}[leftmargin = *, align = left]
    \item $l_x$ ($\in \mathbb{R}_{>0}$), the spatial linking length that is used to find root-level haloes as discussed in Sec. \ref{subsec:rootclusters}. May be given by the user although is set to $\infty$ by default -- which makes \clustar\ treat the input data as a root-level halo.
    
    \item $adaptive$ ($\in \{0, 1, 2\}$), a flag that defines the behaviour of the metric used to calculate the Mahalanobis distances between points as discussed in Sec. \ref{subsec:transformation}. May be given by the user although is set to $1$ by default -- which provides consistently robust results as shown in Sec. \ref{sec:infocontent}.
    
    \item $k_\mathrm{den}$ ($\in \mathbb{N}_{\geq7}$), the number of nearest neighbours that are used to calculate the measure of local density for each point as discussed in Sec. \ref{subsec:density}. May be given by the user although is set to $20$ by default -- which provides consistently robust results as shown in Secs. \ref{sec:optimisation} and \ref{sec:infocontent}.
    
    \item $k_{\mathrm{link}}$ ($\in \mathbb{N}_{\geq7\ \land\ \leq k_\mathrm{den}}$), the number of nearest neighbours that are used to densely connect the points in the data as discussed in Sec. \ref{subsec:aggregation}. May be given by the user although is automatically calculated to be optimal by default as in Sec. \ref{subsec:klinkvalues} -- these optimal values are based on the input data and the value of $k_{\mathrm{den}}$ and ensure maximal cluster completeness and inter-neighbourhood connectivity.
    
    \item $\rho_{\mathrm{threshold}}$ ($\in \mathbb{R}_{\geq1}$), the factor by which the median density of a cluster must be denser than that cluster's surrounds as discussed in Sec. \ref{subsec:rejection}. May be given by the user although is automatically calculated to be optimal by default as in Sec. \ref{subsec:rhothresholdvalues} -- these optimal values are based on the input data and the value of $k_{\mathrm{den}}$ and ensure that the leaf level of the returned clusters are satellite-like overdensities.
    
    \item $f_{\mathrm{reject}}$ ($\in \mathbb{R}_{\geq0\ \land\ \leq1}$), the maximum fraction of points that can be shared by parent-child clusters in the hierarchy as discussed in Sec. \ref{subsec:rejection}. May be given by the user although is set to $0.9$ by default -- which provides a simple to interpret hierarchy that agrees with analyses in \citet{Oliver2020} as discussed in Sec. \ref{subsec:frejectvalues}.
    
    \item $S_{\mathrm{outlier}}$ ($\in \mathbb{R}_{\geq1}$), the local-outlier-factor that is used to reject outlier points from a candidate cluster as discussed in Sec. \ref{subsec:rejection}. May be given by the user although is set to $2.5$ by default -- which provides a moderate level of outlier removal whilst maintaining good clustering results as is shown in Sec. \ref{subsec:soutliervalues}.
\end{enumerate}
Since optimised values for $4$ of these parameters are found, the user only \textit{needs} to consider choosing values for $l_x$, $adaptive$ ,and $k_{\mathrm{den}}$. The details of the algorithm \clustar\ and how to choose these three parameters is outlined in the subsections of Secs. \ref{sec:clustar} and \ref{sec:optimisation}.

\subsection{Defining Root-Level Clusters} \label{subsec:rootclusters}
One major distinction between the group-finding component of the \clustar\ algorithm and that of the \hoptics\ algorithm is that the implementation of \clustar\ does away with having to perform a radial search about each point. This is the largest factor in why \clustar\ is so much faster than \hoptics\ and is also why \clustar\ does not have an equivalent parameter to that of the \optics\ parameter $\epsilon$. In \optics\ this parameter not only aids in the ordering of points but also defines the maximum reachability distance that a clustered point can have -- thereby prescribing an approximate minimum density that a cluster can have. The functionality of $\epsilon$ that leads to the detection and extraction of clusters can be approximated without its presence (refer to Sec. \ref{subsec:aggregation} for details on this), however the effect of defining a cluster's minimum density requires another separate step.

In \hoptics, the choice of $\epsilon$ is redirected to the choice of $\Delta$ -- the overdensity factor. This is the factor by which a field halo (root-level cluster) is denser than the critical density of the Universe, $\rho_{\mathrm{crit}}$. This is accomplished via a mapping between the \code{FOF} linking length, $l_x$, and $\epsilon$ and is used to approximate the \code{FOF} field haloes. In \clustar 's implementation, we offer the functionality of directly computing the \code{3DFOF} field haloes from the positions\footnote{If this functionality is used, then the first $3$ features of the input data must be the Cartesian spatial coordinates of the points.} of the data set as a whole. This splits the input data into some number of field haloes and opens the opportunity for trivial parallelisation over each of these haloes for the remaining steps.

These \code{3DFOF} field haloes will be found by \clustar\ whenever $l_x \neq \infty$. A common choice for $l_x$ when finding field haloes from cosmological simulations is $l_x = 0.2 L_\mathrm{box}/N$ -- where $L_\mathrm{box}$ is the side length of the simulation box and $N$ is the number of particles within it -- which corresponds to field halo overdensities of $\gtrsim 100\bar{\rho}$ \citep{Elahi2019}. Alternatively when $l_x = \infty$ (default), \clustar\ will neglect finding field haloes in this way which results in the root-level cluster becoming the entire data set. The choice of whether to find field haloes using the \code{3DFOF} approach or not must be made under consideration of the input data and the intended output. This choice is outlined in the first few nodes of Fig. \ref{fig:rootlevel}.

Other hierarchical astrophysical clustering algorithms such as \code{VELOCIraptor} \citep{Elahi2019} also perform this step in order to define the root-level clusters before then finding the substructure within them. However due to the possibility of not having to perform this step in \clustar, it is also possible for the user to apply some other algorithm to the data -- or to not do so at all -- in order to perform this step prior to using \clustar. By default $l_x$ is set to $\infty$ which forces \clustar\ to treat the input data as a single root-level halo from which the substructure therein is subsequently found.

\begin{figure}
\begin{center}
\begin{tikzpicture}[node distance=1.5cm, every node/.style={fill=white, text centered, font=\sffamily}], align=center]
    \node[align = center] (input)[base]
    {Input parent cluster and parameters.};
    
    \node[align = center] (pca)[base, below of=input]
    {Take a PCA transformation of the data and\\then rescale each component to unit variance.};
    
    \node[align = center] (kden)[base, below of=pca]
    {Find the $k_\mathrm{den}$ nearest neighbours\\for each point from the transformed data.};
    
    \node[align = center] (density)[base, below of=kden]
    {Find the density of each point using Eqs. \ref{eq:densityestimator} \& \ref{eq:kernelfunction}.};
    
    \node[align = center] (aggregate)[base, below of=density]
    {Produce a clustering, $\mathcal{C}$, using the process in Fig. \ref{fig:aggregate}.};
    
    \node(ifadaptive)[text width = 3.675cm, below of=aggregate, xshift = -2.1625cm]
    {if $\mathcal{C} \neq \varnothing\ \land\ adaptive = True$};
    
    \node[align = center] (adaptivechange)[base, below of=aggregate, xshift=2.1625cm, minimum width = 3.675cm]
    {Keep only first level\\of substructure in $\mathcal{C}$};
    
    \node[align = center] (searchsubstructure)[base, below of=adaptivechange, xshift = -0.75cm, minimum width = 5.175cm]
    {For each $c \in \mathcal{C}$, search for substructure\\within it using the process in Fig. \ref{fig:searchsubstructure}.};
    
    \node[align = center] (output)[base, below of=searchsubstructure, xshift = -1.4125cm]
    {Output the hierarchy of substructure.};
    
    \draw[->](input) -- (pca);
    \draw[->](pca) -- (kden);
    \draw[->](kden) -- (density);
    \draw[->](density) -- (aggregate);
    \draw[<-](ifadaptive) -- +(0, 1);
    \draw[dashed, ->](ifadaptive) -- +(0, -2.5);
    \draw[->](ifadaptive) -- (adaptivechange);
    \draw[->](adaptivechange) -- +(0, -1);
    \draw[->](searchsubstructure) -- +(0, -1);
\end{tikzpicture}
\end{center}
\vspace{-8pt}
\caption{An activity chart of the methods concerned with the set up for finding substructure. A transformation of the data is taken, the nearest neighbour lists are found, and then from them the density of each point is calculated. This generalises the approach to clustering the substructure. If such a clustering is found and $adaptive$ parameter is set to $2$, then the process in this activity chart is invoked again on first level of substructure. This occurs iteratively until there are no more significant clusters found, at which point the process produces a cascade of returns of the adaptive hierarchy of substructure. If $adaptive$ is set to $0$ or $1$, the process returns the hierarchy of substructure that has been found using a parent cluster (globally) defined metric.}
\label{fig:searchsubstructure}
\end{figure}
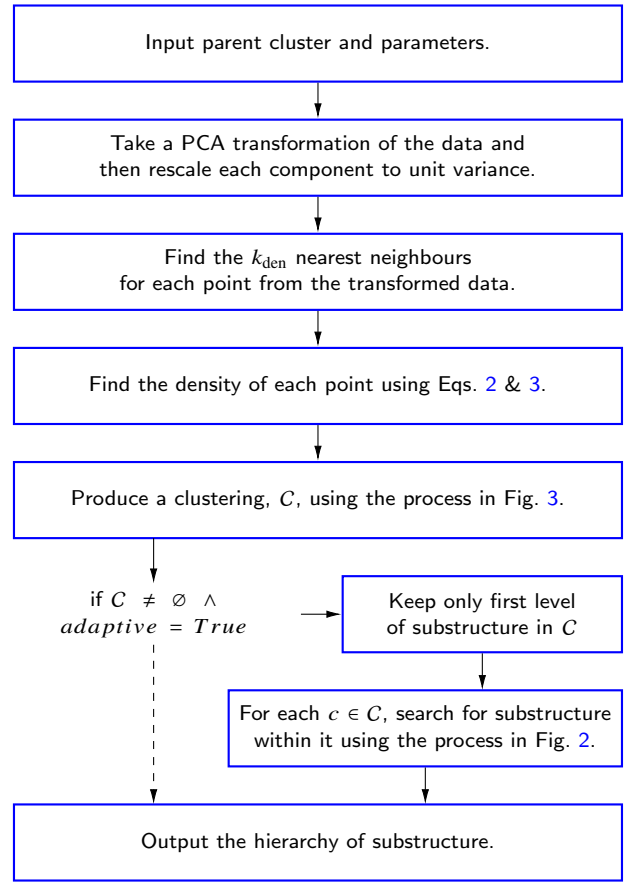

Once the root-level cluster(s) have been found, we add them to a list of unsearched clusters in order to find the substructure within them. Each of the unsearched clusters are independent of each other, meaning that we can search for substructure in parallel. This remains true even when we set $adaptive = 2$, which ensures that \clustar\ iteratively searches within the top level of substructure until there are no more clusters found -- as shown in Figs. \ref{fig:rootlevel}, \ref{fig:searchsubstructure}, and \ref{fig:aggregate}. In this way, setting $adaptive = 2$ creates a locally adaptive metric since the data is transformed before each instance of finding substructure. The following subsection provides details of how these transformations are performed.

\subsection{Data Transformation} \label{subsec:transformation}
When searching for substructure within a previously unsearched cluster, we must implement a strategy that scales the various dimensions so that the clusters found are not overly dependent on a subset of the features. Such a strategy needs to be robust against the unit choice and coordinate orientation. So that we produce the desired effects, we choose to use a Principle Component Analysis (PCA) to first transform the data within the cluster. We then scale each PCA component to unit variance. If we calculate distances via the Euclidean distance metric on the transformed space, we are then effectively using a Mahalanobis distance metric \citep{Mahalanobis1936} such that
\begin{equation} \label{eq:mahalanobismetric}
    s^2(x_i, x_j) = (x_i - x_j)^T \Sigma^{-1}(x_i - x_j),
\end{equation}
where $\Sigma = E[({\bf X} - \mu_{\bf X})({\bf X} - \mu_{\bf X})^T]$ is the covariance matrix of the cluster before the transformation. This guarantees that substructure found within a cluster will be dense with regards to the cluster. When $adaptive = 2$, \clustar\ approximates a locally adaptive Mahalanobis metric since a new covariance matrix is redefined at every level of the hierarchy\footnote{It should be noted that this is not equivalent to the locally adaptive Mahalanobis metric of \code{EnLink} \citep{Sharma2009}, which is constructed via a entropy-based binary space partitioning algorithm and defines a unique covariance matrix for each point a priori to the clustering of the data. While this method is very effective for defining a locally adaptive metric based on the local distribution of points, it has a substantially larger run-time than the top-down method we implement in \clustar.}. Hence the choice of whether to set $adaptive$ to be $0$, $1$ or $2$ must be made in consideration of the trade-off between run-time constraints and the additional clustering power that an adaptive metric gives over that of a global metric.

\subsection{Density Estimation} \label{subsec:density}
Following the transformation of the data, we now seek to find a measure of the local density surrounding each point. We do this by first finding the $k_{\mathrm{den}}$ nearest neighbours for each point within the transformed space. The density for each point is then approximated using the neighbour list and a multivariate kernel within a balloon estimator such that
\begin{equation} \label{eq:densityestimator}
    \rho_i \propto \frac{1}{h_i^d} \sum_{j=1}^{k_{\mathrm{den}}} K\left(\frac{s(x_i, x_j)}{h_i}\right).
\end{equation}
Here $s(x_i, x_j)$ is defined in Eq. \ref{eq:mahalanobismetric}, $d$ is the dimensionality of the feature space, $K$ is a kernel function, and $h_i$ is a smoothing length corresponding to point $p_i$ -- which we choose to be the distance from $p_i$ to it's ${k_{\mathrm{den}}}^{\mathrm{th}}$-most nearest neighbour. Choosing the smoothing length in this way is not atypical \citep[e.g.][]{Sain2002}, however this also has the added bonus of being equal to the core distance from \optics. Together with an appropriate kernel function, $K$, this choice allows \clustar\ to compute a determinable and smoothed analogue of the rigid density estimator that is the reachability distance. For this we implement an Epanechnikov kernel \citep{Epanechnikov1969}, which is defined as
\begin{equation} \label{eq:kernelfunction}
    K(u) \propto (1 - u^2),
\end{equation}
and is $0$ for $u > 1$. Typically there is a normalisation constant that is defined such that the integral of the kernel over the space is $1$. However, as we clarify in the following subsection, the process of determining significant substructure with \clustar\ only relies on density by processing points in order of descending density and by comparing the relative densities of points. For this reason, we do not need to compute the constant factors of Eqs. \ref{eq:densityestimator} and \ref{eq:kernelfunction}.

\subsection{Densely Connecting Points via Neighbourly Aggregation} \label{subsec:aggregation}
Once the data has been transformed, the nearest neighbour lists found, and the local densities computed for any given unsearched cluster, the algorithm now seeks to find the substructure of that cluster by aggregating the points that are densely connected. The key to creating a similar output to that of \hoptics\ without performing any radial search is to densely connect points via their nearest neighbour lists. The \optics\ process allows for the algorithm to gain knowledge about the locality of points nearby to (and from the perspective of) the points that have already been appended to the ordered list. It uses this knowledge to constantly be seeking out regions of higher denser since the next-to-be-ordered point is always the point with the smallest reachability distance i.e. largest density.

\clustar\ mimics this process by further restricting the nearest neighbour lists of each point to the nearest $k_{\mathrm{link}}$ points\footnote{The reason for doing this is that the number of neighbours needed to densely connect points is not necessarily equal to number of neighbours that should be used to calculate the local density of a point. In fact, by restricting the nearest neighbour lists, \clustar\ is able to return a greater resolution on the clustering structure -- as is described in Sec. \ref{subsec:aggregation}.}, finding all points that are local density maxima with respect to their surrounds, and then using these points as seeds from which to densely connect other points to. Seeding the list of dense connections with each of the local maxima ensures that the aggregation process within \clustar\ does not need to seek out regions with a higher density than the points that have already been connected together. Instead it simply connects each point (that isn't a local density maximum) in order of decreasing density to a set of already densely connected points -- this process is similar to those of the core-search in \code{VELOCIraptor} \citep{Elahi2019} as well as the group-finders of \code{SUBFIND} \citep{Springel2001} and \code{EnLink} \citep{Sharma2009}.

\begin{figure}
\begin{center}
\begin{tikzpicture}[node distance=1.5cm, every node/.style={fill=white, text centered, font=\sffamily}], align=center]
    \node[align = center] (input)[base]
    {Input the points ($P$), $k_\mathrm{den}$ nearest\\neighbours, densities, and parameters.};
    
    \node[align = center] (reduceneighbours)[base, below of=input]
    {Construct the $k_\mathrm{link}\ (\leq k_\mathrm{den})$ nearest neighbours\\for each point from their $k_\mathrm{den}$ nearest neighbours.};
    
    \node[align = center] (localmaxima)[base, below of=reduceneighbours]
    {Find the set of local density maxima, $M$, and seed the lists\\of densely connected points with them, $D = \{\{p\} \mid \forall p \in M\}$.};
    
    \node[align = center] (dcinit)[base, below of=localmaxima]
    {Initialise the sets unprocessed and clustered points,\\$U = P\setminus M$ and $\mathcal{C} = \varnothing$ respectively.};
    
    \node[align = center] (nextpoint)[repeat1, below of=dcinit, xshift = 0.5cm]
    {Retrieve the next point, $p_i$, where\\$i = \operatorname*{arg\,max}_i \{\rho_i \in \rho \mid p_i \in U\}$. Remove $p_i$ from $U$.};
    
    \node[align = center] (connect)[repeat1, below of=nextpoint]
    {Find the set of neighbourly connections,\\$T = \{d \in D \mid \exists p_j \in N_i'\setminus U,\ p_j \in d\}$.};
    
    \node(ifmulticonnect)[text width = 1.2cm, below of=connect, xshift=-1.5cm]
    {if $|T| > 1$};
    
    \node[align = center] (oldcluster)[repeat1, below of=ifmulticonnect, minimum width = 3cm]
    {Append $p_i$ to $d \in D$,\\where $d \in T$.};
    
    \node[align = center] (removeconnections)[repeat2, below of=connect, xshift=2cm, minimum width = 3cm]
    {Remove all $d \in T$\\from $D$.};
    
    \node[align = center] (newcluster)[repeat2, below of=removeconnections, minimum width = 3cm]
    {Append the new set\\ $\bigcup T \cup \{p_i\}$ to $D$.};
    
    \node[align = center] (cleanrho)[repeat2, below of=newcluster, minimum width = 3cm]
    {Filter $T$ using\\Eqs. \ref{eq:rhocondition} and \ref{eq:singleleaf}.};
    
    \node[align = center] (appendtoc)[repeat2, below of=oldcluster, minimum width = 3cm]
    {Append the elements\\of $T$ to $\mathcal{C}$.};
    
    \node[align = center] (whileu)[repeat1, left of=oldcluster, xshift = -1.0cm, yshift = -0.75cm, rotate=90, minimum width = 2.5cm]
    {While $U \neq \varnothing$};
    
    \node[align = center] (cleanrhoagain)[repeat2, below of=appendtoc, minimum width = 3cm]
    {Filter $D$ using\\Eqs. \ref{eq:rhocondition} and \ref{eq:singleleaf}.};
    
    \node(ifmerged)[text width = 1.21cm, left of=cleanrhoagain, xshift=-1.0cm]
    {if $|D| > 1$};
    
    \node[align = center] (merge)[repeat2, below of=cleanrho, minimum width = 3cm]
    {Append the elements\\of $D$ to $\mathcal{C}$.};
    
    \node[align = center] (cleanfreject)[base, below of=merge, xshift = -2.5cm]
    {Filter $\mathcal{C}$ using Eq. \ref{eq:fcondition}.};
    
    \node[align = center] (cleansoutlier)[base, below of=cleanfreject]
    {Remove any $p$ from any $c$ in $\mathcal{C}$ not satisfying Eq. \ref{eq:rhocut}.};
    
    \node[align = center] (output)[base, below of=cleansoutlier]
    {Output $\mathcal{C}$, the subclusters of $P$.};
    
    \draw[->](input) -- (reduceneighbours);
    \draw[->](reduceneighbours) -- (localmaxima);
    \draw[->](localmaxima) -- (dcinit);
    \draw[<-](nextpoint) -- +(0, 1);
    \draw[->](nextpoint) -- (connect);
    \draw[<-](ifmulticonnect) -- +(0, 1);
    \draw[dashed, ->](ifmulticonnect) -- (oldcluster);
    \draw[->](ifmulticonnect) -- (removeconnections);
    \draw[->](removeconnections) -- (newcluster);
    \draw[->](newcluster) -- +(0, -1);
    \draw[->](oldcluster.west) -- +(-0.5, 0);
    \draw[->](cleanrho) -- (appendtoc);
    \draw[->](appendtoc.west) -- +(-0.5, 0);
    \draw[->](whileu) |- (nextpoint);
    \draw[dashed, ->](whileu) -- (ifmerged);
    \draw[dashed, ->](ifmerged) -- +(0, -1);
    \draw[->](ifmerged) -- (cleanrhoagain);
    \draw[->](cleanrhoagain) |- (merge);
    \draw[->](merge) -- +(0, -1);
    \draw[->](cleanfreject) -- (cleansoutlier);
    \draw[->](cleansoutlier) -- (output);
\end{tikzpicture}
\end{center}
\vspace{-4pt}
\caption{The activity chart for the aggregation and rejection process that constructs a hierarchy of subclusters similar to those from \hoptics. The nearest neighbour lists are reduced and from them the local density maxima are found. The lists of densely connected points are then seeded with these local maxima. In order of decreasing local density, the points are either appended to an existing list of densely connected points or used to merge multiple into a new one. In the event of the latter and subject to the conditions of Eqs. \ref{eq:rhocondition} and \ref{eq:singleleaf}, these lists are considered to be potential clusters. Following this inner loop, if not all points were densely connected, the remaining connected lists are subject to these same conditions in order to become potential clusters. The hierarchy of clusters is then cleaned using Eq. \ref{eq:fcondition}, and each remaining cluster is also cleaned subject to Eq. \ref{eq:rhocut}.}
\label{fig:aggregate}
\end{figure}
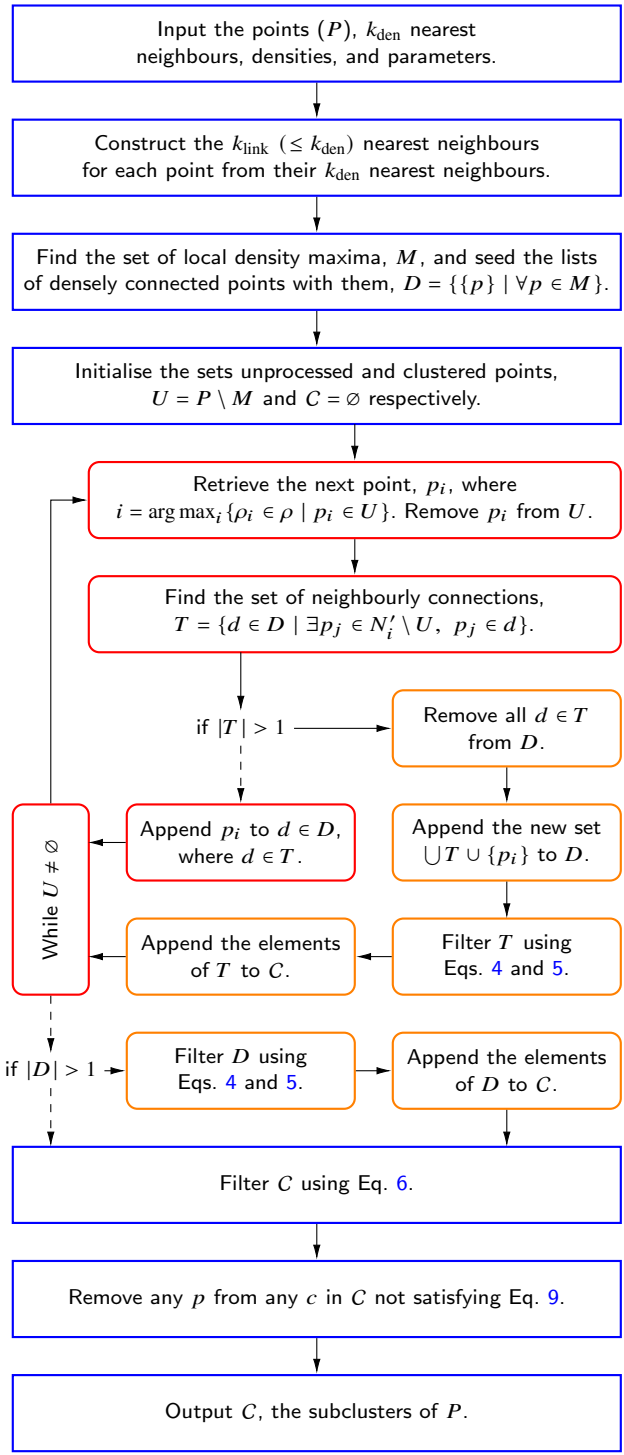

The key condition that \clustar\ exploits here is that even though the \optics\ algorithm gains knowledge of all unordered points within a radius of $\epsilon$, the reachability distance of those points will typically be large if they are not directly connectable via the $N_{\mathrm{min}}$ nearest neighbours of either itself or the already ordered points. Hence at any given time during \optics\ ordering process, the next-to-be-ordered point is almost always a neighbour of a point (or vice versa) that has already been ordered. The only time that it isn't as such, is if the next-to-be-ordered point cannot be directly connected through the $N_{\mathrm{min}}$-sized neighbourhoods of those points that have already been ordered -- or of such a neighbourhood of it's own. This does not mean that such a point can never be connected through $N_{\mathrm{min}}$-sized neighbourhoods, however doing so will require traversing points that are less dense than itself -- densely connecting points concurrently. Such a scenario where the point with the smallest reachability distance cannot be densely connected through $N_{\mathrm{min}}$-sized neighbourhoods is possible in \optics, although whether two points belonging to entirely separate sets of densely connected points are at all associated is somewhat uncertain in an astrophysical context. This distinction between \clustar\ and \hoptics\ is investigated further in Sec. \ref{sec:comparison}.

The process of aggregating points is iteratively performed; by retrieving the unprocessed point with the highest density ($p_i$), finding the set(s) of densely connected points ($T$) that have members within the $k_{\mathrm{link}}$ nearest neighbours of $p_i$, then either appending $p_i$ to an existing list of densely connected points or constructing clusters from $T$ and merging the lists of $T$ to create a new list of densely connected points (with $p_i$ included). The condition for whether clusters are to be constructed and a new list of densely connected points created, is if $|T| > 1$ i.e. if the point $p_i$ can be densely connected to multiple sets of densely connected points. In this scenario, the previous sets of $T$ are only retained as potential clusters if they satisfy Eq. \ref{eq:rhocondition} -- a relative density condition dependent upon the trainable parameter $\rho_{\mathrm{threshold}}$. More details on this condition and others that are responsible for cluster rejection are given in Sec. \ref{subsec:rejection}.

The circumstances under which $|T| > 1$ for any given point depend strongly upon the value of $k_{\mathrm{link}}$ since this determines how many neighbours are considered for connecting the points. Decreasing $k_{\mathrm{link}}$ ensures that a greater number of points can be densely connected to the potential clusters before eventually finding a connecting point whose neighbourhood satisfies $|T| > 1$. In effect, this gives a greater resolution to the clusters and, since $k_{\mathrm{link}}$ can be much less than $k_{\mathrm{den}}$, is typically greater even than those from \hoptics. Of course, there is a reasonable lower limit to the value of $k_{\mathrm{link}}$ that can be found by ensuring that all points be densely connected under a uniform distribution of points. This lower limit is investigated further in Sec. \ref{subsec:klinkvalues}. Finally, if all points within the previously unsearched parent cluster do not become densely connected through this aggregation process and if more than $1$ of the densely connected subsets have at least $k_{\mathrm{link}}$ points, then we append those that do to the list of substructures as well.

\subsection{Cluster and Cluster Member Rejection} \label{subsec:rejection}
The cluster extraction algorithm of \hoptics\ is implemented in \clustar, however it is not entirely its own separate process that simply takes place after the aggregation/ordering of points as it is in \hoptics. To detect and extract clusters from the reachability plot, \hoptics\ performs a set of routines -- each of which are described in Sec. 3.2 of \citet{Oliver2020}. First, the algorithm determines all local maxima of the reachability plot from \optics\ and then using this, builds the hierarchy of clusters by taking contiguous subsets of the ordered list that are situated on either side of each local maxima -- ensuring that each subset only contains points with reachability distances less than that at the corresponding local maxima. \clustar\ performs these steps by ensuring that points are aggregated in order of decreasing local density and then by considering lists of densely connected points that are connected to one another via the neighbourhood of a less dense point as potential clusters.

\hoptics\ then rejects all potential clusters that contain less than $N_{\mathrm{min}}$ points or that have median local densities less than $\rho_{\mathrm{threshold}}$ times the surrounding density of those potential clusters. Following this, the algorithm rejects all single leaf potential clusters. These criteria are fulfilled by \clustar\ in a single step during the aggregation process when multiple lists of densely connected points are considered to be potential clusters. As such, \clustar\ conducts the following:
\begin{align} \label{eq:rhocondition}
    \mathrm{reject\ any}\ d\ \mathrm{from}\ T\ \mathrm{if}&\ \mathrm{it\ has\ no\ children\ and\ either;}\notag\\
    &|d|<k_{\mathrm{link}}\ \mathrm{or}\notag\\
    &\mathrm{median} \{\rho_j \mid p_j \in d \}/\rho_i < \rho_{\mathrm{threshold}}.
\end{align}
Here, $\rho_i$ is the local density of $p_i$ -- the point responsible for densely connecting the otherwise disconnected sets in $T$. Following the rejection of all $d \in T$ that do not meet these criteria, the set $T$ is then subject to the condition that
\begin{equation} \label{eq:singleleaf}
    \mathrm{if}\ |T| = 1,\ \mathrm{set}\ T = \varnothing.
\end{equation}
The conditions in Eqs. \ref{eq:rhocondition} and \ref{eq:singleleaf} are related to steps $3$ and $4$ from the cluster extraction process of \hoptics. Next, for all parent-child cluster pairs sharing at least $f_{\mathrm{reject}}$ of the parent’s points; \clustar\ rejects the child if it has child clusters of its own, otherwise \clustar\ rejects the parent. To re-iterate, given a set of substructure $S$ and any parent-child pair $s_{\mathrm{parent}}$ and $s_{\mathrm{child}}$,
\begin{align} \label{eq:fcondition}
    \mathrm{if}\ |s_\mathrm{child}|/|s_\mathrm{parent}|&\geq f_{\mathrm{reject}}\ \mathrm{then},\notag\\
    &\mathrm{reject}\ s_\mathrm{child}\ \mathrm{if;}\ \exists s \in S\ \mathrm{such\ that}\ s \subset s_{\mathrm{child}},\notag\\
    &\mathrm{else;\ reject}\ s_\mathrm{parent}.
\end{align}
This is equivalent to step $5$ of the extraction process in \hoptics. Steps $6$ and $7$ of the \hoptics\ extraction process are concerned with removing outliers based on how their local density compares with the rest of their neighbours. While this approach works in the context of using a reachability distance as the measure of inverse density, it can be improved upon with a more robust measure of density such as that that \clustar\ computes. The measure used by \hoptics\ is the local-outlier-factor which relies upon the local-reachability-density \citep{Breunig1999}, both of which are defined in Eqs $3$ and $4$ of \citet{Oliver2020}. In effect, the local-outlier-factor compares how reachable a point is from its neighbours to how reachable its neighbours from their own. We provide \clustar\ with a comparable measure with which to define outliers that incorporates the estimate of local density it uses from Eq. \ref{eq:densityestimator}.

We define the kernel density estimate analogue of the local-reachability-density of a point $p_i$ with respect to transformed space as
\begin{equation} \label{eq:lrd}
    \mathrm{lrd}(p_i) \propto \frac{1}{\sum\limits_{p_j \in N_{k_{\mathrm{den}}}} \mathrm{min}\left(\rho_i,\ \rho_j \right)^{-1/d}},
\end{equation}
where $N_{k_{\mathrm{den}}}$ is the $k_\mathrm{den}$ nearest neighbours of $p_i$ within the previously unsearched cluster. \clustar\ then uses this to similarly compute an analogue of the local-outlier-factor,
\begin{equation} \label{eq:lof}
    \mathrm{lof}(p_i) = \frac{\sum\limits_{p_j \in N_{k_{\mathrm{den}}}} \frac{\mathrm{lrd}(p_j)}{\mathrm{lrd}(p_i)}}{k_\mathrm{den}}.
\end{equation}
In \hoptics, a point is considered an outlier from a cluster if the local-outlier-factor (calculated with respect to that cluster) is larger than $S_{\mathrm{outlier}}$. In \clustar\ we adjust this rule by first defining a cut-off density,
\begin{equation} \label{eq:rhocut}
    \rho_\mathrm{cut} = \mathrm{min}\{\rho_i \mid \mathrm{lof}(p_i) < S_\mathrm{outlier}\},
\end{equation}
which we then use to reject all points from a cluster that have a density less than this. This ensures that only local-outliers at the outskirts of a cluster are rejected and not those that are embedded deeper within it -- a more appropriate regime for outlier detection and rejection within a hierarchical set of clusters.

\section{Synthetic Data} \label{sec:syntheticdata}
Within the remainder of this paper we compare (Sec. \ref{sec:comparison}), train (Sec. \ref{sec:optimisation}), and scrutinise (Sec. \ref{sec:infocontent}) our algorithm against synthetic survey data of Milky Way (MW) type galaxies produced by \code{Galaxia} \citep{Sharma2011}. Given certain survey restrictions, such as one or more colour-magnitude bounds, a survey size, and geometry, \code{Galaxia} is able to return a synthetic catalogue of stars in accordance with a given model of the MW. The code is generalised enough so that it can also accept star formation rates, age–metallicity relations, age–velocity-dispersion relations, and analytic density distribution functions from which it can also use to produce stellar catalogues. 

Among the synthetic data that can be produced with \code{Galaxia} are resamplings of the $11$ $\Lambda$CDM stellar haloes from \citet{Bullock2005} and the complementary $6$ artificial stellar haloes from \citet{Johnston2008}. The original $\Lambda$CDM haloes are simulated using a hybrid semi-analytical plus hydrodynamic $N$-body approach that replicates a density profile and satellite distribution that is similar to the MW. The simulation model assumes a $\Lambda$CDM cosmology with parameters $\Omega_m = 0.3$, $\Omega_\Lambda = 0.7$, $\Omega_bh^2 = 0.024$, $h = 0.7$, and $\sigma_8 = 0.9$. The simulations generate, track, and evolve a number of individual satellites that are first modelled as $N$-body dark matter systems within a parent galaxy whose disk, bulge, and halo are represented by time dependent semi-analytic functions. Semi-analytical prescriptions are then used to assign star formation histories and leaky accreting boxes to each. A chemical enrichment model is also used to calculate the metallicities as a function of age for the stellar populations \citep{Robertson2005, Font2006}. Ultimately, the dark matter distributions of the satellites follow NFW profiles \citep{Navarro1996} while the stellar distributions follow King profiles \citep{King1962} -- the latter of which is constructed in order to reproduce an agreement with the structural properties of the Local Group's dwarf galaxies.

Each satellite created in this way has three main model parameters; the time since accretion ($t_{\mathrm{acc}}$), the luminosity ($L_{\mathrm{sat}}$), and the orbital circularity ($\epsilon = J/J_{\mathrm{circ}}$). The distribution of these parameters specify the accretion history of a halo and as such a further $6$ artificial haloes were created in \citet{Johnston2008} for the purpose of studying the effects of different accretion histories on the properties of haloes. These $6$ artificial haloes have accretion events that are predominantly; radial ($\epsilon < 0.2$); circular ($\epsilon > 0.7$); old ($t_{\mathrm{acc}} > 11$ Gyr); young ($t_{\mathrm{acc}} < 8$ Gyr); high luminosity ($L_{\mathrm{sat}} > 10^7 L_{\mathrm{\odot}}$); and low luminosity ($L_{\mathrm{sat}} < 10^7 L_{\mathrm{\odot}}$). Importantly, the output from \code{Galaxia} also bestows each star with a label that corresponds to which satellite group it belonged to at the time that that satellite was created within the simulation -- this allows us to test how well \clustar\ performs in Sec. \ref{sec:optimisation}. In addition to this, \code{Galaxia} maintains a list of satellite properties that includes information on whether the satellite is self-bound or not. More details on these simulations can be found in Sec. 3.4 of \citet{Sharma2011} and references therein.

We use \code{Galaxia} to produce a random sample from each synthetic galaxy in both sets of the stellar haloes. From these we use various combinations of the spatial (denoted by ${\bf x} = (x, y, z)$ in the later sections), kinematic (denoted by ${\bf v} = (v_x, v_y, v_z)$ in the later sections), and chemical (denoted by ${\bf m} = ($[Fe/H]$, $[$\alpha$/Fe]$)$ in the later sections) information in order to compare, optimise, and apply our algorithm. For each galaxy, the points are contained inside the $282$ kpc virial radius of the corresponding host dark matter halo -- these have an overdensity factor of $\Delta \approx 337$ times the mean dark matter density of the universe. As such we treat these galaxies as field haloes and hence do not use \clustar\ to find field haloes before finding substructure. Since these field haloes and those that are found via 3D \code{FOF} are well defined, the analysis in the following sections is an assessment of the substructure finding component of \clustar.


\begin{figure*}
    \centering
    \includegraphics[trim={20mm 0mm 20mm 8mm}, clip, width=\textwidth]{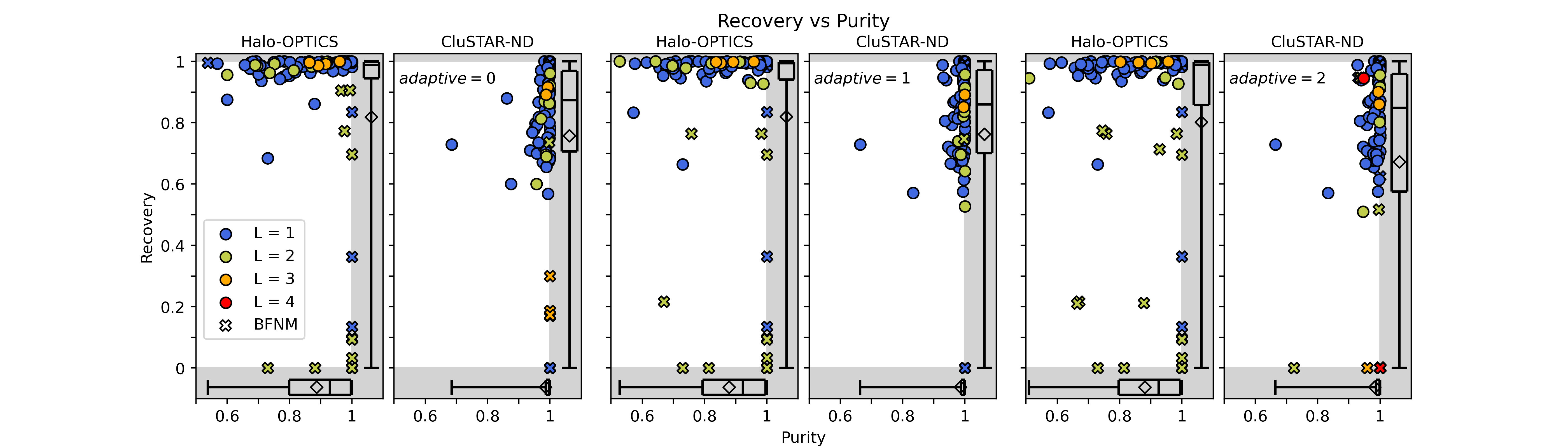}
    \vspace{-0.5cm}
    \caption{A clustering comparison of the outputs of \hoptics\ and \clustar\ after having been applied to the 3D spatial dimensions of the synthetic galaxies from \code{Galaxia}. Each panel belongs to one of three pairs and each pair corresponds to a different comparison wherein \clustar\ has been used with a different metric adaptivity setting -- which is annotated on the relevant panels. Plotted within each panel are the recovery and purity fractions of each best-fitting cluster to either; the \hoptics\ predicted clusters (left panel in each pair) or the \clustar\ predicted clusters (right panel in each pair). Best-fitting clusters are those that produce the maximum Jaccard index in Eq. \ref{eq:maxjaccardindex} (i.e. $T^* = \operatorname*{arg\,max}_{T \in S_2} J(C, T)$), the recovery and purity are then found according to Eq. \ref{eq:recoveryandpurity}. To clarify, the label at the top of each panel indicates which clustering is used as $S_1$ within these equations. Every cluster, except for the root-level clusters (L = $0$), appear in these panels as a marker coloured by its level within the hierarchy and shaped by whether or not its best-fitting cluster is a mutually best-fitting -- note that {\it BFNM} here stands for {\it best-fitting not mutual}. The axes are extended beyond the possible recovery and purity values for easier readability, this extended area is coloured grey. Within this area on each panel and for both the recovery and purity distributions, there is a box and whisker plot that denotes the $Q_0$ -- $Q_4$ quartiles with mean values also indicated with diamonds markers.}
    \label{fig:outputcomparison}
\end{figure*}

\section{A Comparison with \textsc{Halo-OPTICS}} \label{sec:comparison}
Our first assessment of \clustar's performance is done by comparing it to \hoptics. We compare both the clustering and the run-times of these algorithms when applied to the synthetic haloes outlined in Sec. \ref{sec:syntheticdata}. Since \hoptics\ is designed for clustering on 3D spatial data, we only apply these algorithms to the 3D positions of the points in these haloes.

We apply \clustar\ in a way that treats each synthetic galaxy as a field halo ($l_x = \infty$) -- due to each galaxy only containing points within its virial radius. Similarly, for \hoptics\ we find the smallest value of $\epsilon$ that will guarantee that every point in these reduced galaxies is a part of at least one core-point's $N_\mathrm{min}$ nearest neighbours -- thereby ensuring that every point will be given a reachability distance less than or equal to $\epsilon$. By applying these algorithms to ready-made galactic data sets, we can ensure that we are only comparing the substructure found and not the root-level clusters -- which in both cases are equal to the entire data set due to the above.

In order to assess how adequately \clustar\ reproduces \hoptics, we set $k_\mathrm{den} = 20$ and similarly for \hoptics\ we set $N_\mathrm{min} = 20$. For \clustar\ we set $k_\mathrm{link} = 7$, as is described in Sec. \ref{subsec:klinkvalues}. For both codes we use the near-optimal parameters found in \citet{Oliver2020} such that $\rho_\mathrm{threshold} = 2$, $f_\mathrm{reject} = 0.9$. However since the outlier detection is different between the codes, we set $S_\mathrm{outlier} = \infty$ which ensures that there are no outliers removed from clusters. We review the optimal choice of these parameters in Secs. \ref{subsec:rhothresholdvalues} -- \ref{subsec:frejectvalues}.

\subsection{Output Similarity} \label{subsec:outputsimilarity} 
To appropriately compare the output of the two codes, we find the best-fitting match from the \hoptics\ catalogue of clusters for each of the \clustar\ clusters (ignoring the root-level clusters which are fixed to be equal). As a measure of which pair is {\it best-fitting} we use the maximum Jaccard index \citep{Jaccard1912}, which -- given two clusterings of the same data set ($S_1$ and $S_2$) -- allows us to not only find which clusters from $S_1$ are best-fitting to the clusters from $S_2$ but to also compare how well the clusters in $S_2$ are matched by those in $S_1$. Hence for two such clusterings the maximum Jaccard index is given by
\begin{align} \label{eq:maxjaccardindex}
    &J_\mathrm{max}(C) = \mathrm{max}\{J(C, T) \mid \forall T \in S_2\},\ C \in S_1,\ \mathrm{where}\notag\\
    &J(C, T) = \frac{|C\cap T|}{|C\cup T|}.
\end{align}
Once each cluster from $S_1$ has been assigned a best-fitting cluster from $S_2$, we then compute the recovery and purity fractions of the pairs. We define these such that
\begin{align} \label{eq:recoveryandpurity}
    &R(C) = \frac{|C\cap T^*|}{|T^*|},\ \mathrm{and} \notag\\
    &P(C) = \frac{|C\cap T^*|}{|C|}.
\end{align}
where $T^* \in S_2$ is the best-fitting match to $C \in S_1$ -- i.e. $T^* = \operatorname*{arg\,max}_{T \in S_2} J(C, T)$. Neatly, this implies that
\begin{align} \label{eq:jacclowerbound}
    &\frac{1}{J_\mathrm{max}(C)} = \frac{1}{R(C)} + \frac{1}{P(C)} - 1,\ \mathrm{and\ hence} \notag\\
    &J_\mathrm{max}(C) \leq \mathrm{min}\{R(C),\ P(C)\}.
\end{align}

Fig. \ref{fig:outputcomparison} depicts the recovery and purity fractions determined in this way for all of the best-fitting cluster pairs found within the synthetic haloes from \code{Galaxia} for various values of the $adaptive$ parameter. Here we see that the clusterings from \clustar\ are best-fitted by those from \hoptics\ in a way that most commonly leads to clusters from \clustar\ being encompassed by those from \hoptics\ -- indicated in each comparison by high purity and varied recovery in the left panel of each pair and the opposite in the right panel of each pair. This matches the trends seen in both Figures 6 and 10 of \citet{Oliver2020} where \hoptics\ was typically shown over-encompassing both mock cluster sets and predicted clusters found by \code{VELOCIraptor} \citep{Elahi2019}. Regardless, this property of the best-fitting clusters from \clustar\ and \hoptics\ suggests that \clustar\ may benefit from an additional method to allocate points to the clusters that it already finds. Such a task could be achieved via an expectation-maximisation-like technique that allocates previously noisy points to already existing clusters from \clustar\ depending on how well they would assimilate within them.

A small number of clusters are also shown in Fig. \ref{fig:outputcomparison} to have high purity and very low recovery ($< 0.2$). These clusters are those that have been found by one algorithm but not the other -- hence they are not mutually best-fitting -- and are consequences of: 1) \clustar's density estimate being more reliable than that of \hoptics; 2) \clustar\ being able to robustly detect clusters with a smaller number of points; and 3) the algorithmic differences between connecting points. We do also see a few high recovery and high purity clusters whose best-fitting cluster is not mutually best-fitting returned by both codes. These are clusters from one code that effectively sit between two levels of the hierarchy in the other code -- this can produce a good fit to a cluster that is not necessarily a mutually best-fitting cluster. These clusters arise for the same reasons as above.

Overall, it is clear that \clustar\ is able to reproduce the clusters and hierarchy of \hoptics\ with a similar clustering power. We see from Fig. \ref{fig:outputcomparison} and the box and whisker plots therein that typically the clusters from \clustar\ (\hoptics) are mutually best-fitting to those from \hoptics\ (\clustar) with both high recovery and high purity with medians of $\sim0.860$ ($\sim0.990$) and $\sim0.998$ ($\sim0.926$) respectively.

For completeness, we also discuss the clustering power of these outputs from \hoptics\ in comparison with the optimised outputs of \clustar\ in Sec. \ref{sec:infocontent}. In Sec. \ref{sec:infocontent} we see that overall \clustar\ is expected to provide an equal clustering power to \hoptics, however the clustering power does tend to differ slightly on a per galaxy basis. This difference is due to the competing effects of \clustar\ having a more robust density estimation, while the ordering process of \hoptics\ can more easily gather like-satellite points before joining child clusters into their shared parent cluster. We will now define this measure of clustering power as it is used throughout both Secs. \ref{sec:optimisation} and \ref{sec:infocontent}.

\subsubsection{Measuring Clustering Power} \label{subsubsec:clusteringpower}
In order to measure the clustering power of \clustar\ we must construct a scalar function that reduces the complexity of comparing the quality of fit between an entire \clustar\ clustering output when applied to a synthetic galaxy and the ground truth labels of that synthetic galaxy. Since each galaxy has been constructed through the accretion of satellites (refer to Sec. \ref{sec:syntheticdata} for details), the ground truth labels form a flat clustering without noise, i.e. each data point in the galaxy is a member of exactly one satellite. This is a fundamentally distinct clustering type from the hierarchical clustering with noise that \clustar\ produces -- so we only find the goodness of fit between the leaf clusters produced by \clustar\ and the satellite labels within each galaxy.

In order to compare a leaf-clustering produced by \clustar\ ($C_g$) to the ground truth clustering ($T_g$) defined over the same galaxy ($g$), we construct a mutual-information-based objective function similar to that built by \citet{Vinh2009}. To do this we must first append an additional cluster to those in $C_g$ to create $C_g^*$ so that the sets $C_g^*$ and $T_g$ are defined over the same set of data points. The additional cluster is made of the remaining points in the data set that have not been clustered into the leaf clusters returned by \clustar. This means that $C_g^*$ is an artificial construction of a flat clustering without noise using $C_g$ which we can then easily match to the flat clustering without noise that is $T_g$.

The mutual-information between $C_g^*$ and $T_g$, $I(C_g^*; T_g)$, is then the amount of information obtained about the true clusters by having observed the predicted clusters \citep{Shannon1948}.
\begin{align} \label{eq:mutualinformation}
    &I(X;Y) = H(X) - H(Y|X),\ \mathrm{where} \notag\\
    &H(X) \equiv -\sum_{x \in X} P(x)\ \mathrm{log}(P(x)),\ \mathrm{and} \notag\\
    &H(Y|X) \equiv -\sum_{x \in X,\ y \in Y} P(x, y)\ \mathrm{log}\bigg(\frac{P(x, y)}{P(x)}\bigg).
\end{align}
This measure is non-negative and will have a different theoretical maximum depending on the synthetic galaxy in question. For our purposes, we normalise $I(C_g^*; T_g)$ between $0$ and $1$ such that these values represent the absolute worst and best values that a clustering algorithm can be expected to have respectively (see below). We then also take the average of this normalised value over each of the synthetic galaxies so that we may compare sets of \clustar\ clusterings to sets of ground truth clusterings. Let $\mathcal{S}$ represent some feature space combination on which the predicted clusters are dependent, then the resultant objective function we use is defined as
\begin{align} \label{eq:objectivefunction}
    &F(\mathcal{S}) = \frac{1}{N_\mathrm{galaxies}}\sum_{g \in \mathrm{galaxies}} F_g(\mathcal{S}),\ \mathrm{where} \notag\\
    &F_g(\mathcal{S}) = \frac{I(C_g^*(\mathcal{S}); T_g) - E[I(R_g^*(\mathcal{S}); T_g)]}{H(T_g) - E[I(R_g^*(\mathcal{S}); T_g)]}.
\end{align}
Here $H(T_g)$ is the entropy of $T_g$ (this normalises the maximum of $F_g(\mathcal{S})$ to $1$ since $I(T_g; T_g) = H(T_g)$) and $E[I(R_g^*(\mathcal{S}); T_g)]$ is the expected mutual-information that arises when a clustering, $R_g^*(\mathcal{S})$, with equal cluster sizes as $C_g^*(\mathcal{S})$ has been created via random assignment (this normalises the minimum of $F_g(\mathcal{S})$ to $0$ since no clustering algorithm can ever do worse than random assignment). Unlike \citet{Vinh2009} (who relies on a hypergeometric model), we calculate $E[I(R_g^*(\mathcal{S}); T_g)]$ empirically by taking the average of the mutual-information that is found when points have been randomly assigned to a mock clustering, $R_g^*(\mathcal{S})$, with the number of clusters and number of points within each cluster being equal to those in $C_g^*(\mathcal{S})$. This average is calculated using $10$ random realisations of $R_g^*(\mathcal{S})$ which we find to be sufficient since the variance of $I(R_g^*(\mathcal{S}); T_g)$ is small compared to its expectation.

The normalisation adjustments we have made mean that $F_g(\mathcal{S})$ can now be interpreted as the proportion of {\it relevant} information obtained about the true clusters of a galaxy by having observed the predicted leaf clusters produced from \clustar\ when applied to that galaxy. Similarly, $F(\mathcal{S})$ is the average of this proportion across the galaxies. We use these mutual-information-based measures of clustering power throughout both Secs. \ref{sec:optimisation} and \ref{sec:infocontent} as they provide the means to compare \textit{entire} clusterings ($F_g(\mathcal{S})$) and \textit{sets} of entire clusterings ($F(\mathcal{S})$).

While it is possible to construct an objective function with a similar purpose from the easy-to-interpret measures of recovery/purity/Jaccard index, it is non-trivial to reduce these to an appropriate scalar measure of clustering power as these are designed to be used to compare one cluster to another (rather than sets of clusters). Most critically, these measures do not account for true-negative predictions and so optimising the parameters of \clustar\ (as is done throughout Sec. \ref{sec:optimisation} with $F(\mathcal{S})$) by using an objective function built from such measures will likely create an artificial incentive for \clustar\ to make fewer predictions on clusters. These measures also do not appropriately reward \clustar\ for having matched a true cluster with two predicted clusters -- clearly this is not as nice as a perfect match, but it is better than a $J_\mathrm{max}(C) = 0.5$ match for example. Nevertheless, maximising $F(\mathcal{S})$ does also maximise the recovery, purity, and Jaccard index that the predicted clusters will have  -- it just does so without promoting fewer predictions. In fact, a value of $F_g(\mathcal{S}) = f$ is approximately equivalent to having $\sim f\times |T_g|$ true clusters matched perfectly ($J_\mathrm{max}(C) = 1$) by \clustar, of course, this relation is not one-to-one as $F_g(\mathcal{S}) = f$ could also occur via a larger number of predicted clusters with smaller $J_\mathrm{max}(C)$ values -- although, by adjusting $F_g(\mathcal{S})$ for random chance allocation we have limited this trade-off to meaningful predictions.

\subsection{Run-time Disparity} \label{subsec:runtimedisparity}
\begin{figure}
    \centering
    \includegraphics[trim={5mm 5mm 11mm 15mm}, clip, width=\columnwidth]{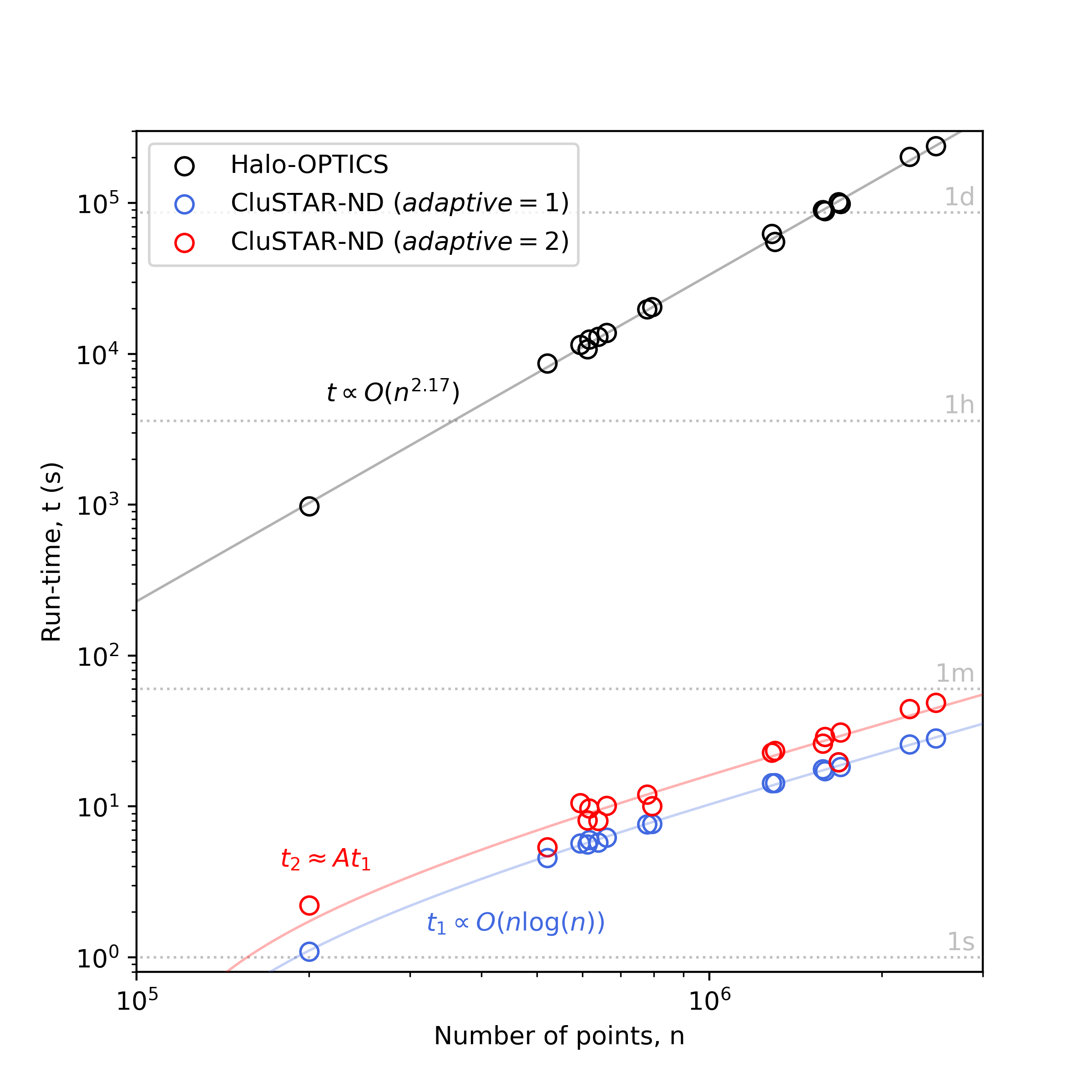}
    \vspace{-0.5cm}
    \caption{The run-times of both \clustar\ and \hoptics\ when applied to the 3D spatial dimensions of the synthetic galaxies from \code{Galaxia}. The lines of best fit correspond to an $O(n\mathrm{log}(n))$ time complexity for \clustar\ and an $O(n^{2.17})$ time complexity for \hoptics. We see that even when applied to the smallest size galaxy ($\sim2\times10^5$ points) in the suite \clustar\ is $3$ orders of magnitude faster than \hoptics, and due to the difference in time complexities, this speed increase balloons outwards to nearly $5$ orders of magnitude when applied to the largest of these galaxies ($\sim2.5\times10^6$ points). While some of this run-time disparity can be attributed to changing clustering structure (see Sec. \ref{subsec:runtimedisparity}), the modest run-time of \clustar\ makes it ideally suited for application to large data sets.}
    \label{fig:runtimescomparison}
\end{figure}

We now compare run-times of \clustar\ and \hoptics\ using synthetic galaxies from \code{Galaxia} in Fig. \ref{fig:runtimescomparison}. All runs were performed using a single core on an Intel i5 vPro processor and both codes are written using \code{Python3} but do make use of optimised numerical packages such as \code{NumPy} \citep{Numpy}, \code{SciPy} \citep{SciPy}, \code{Scikit-learn} \citep{scikit-learn}, and \code{Numba} \citep[\clustar\ only;][]{Numba}. We see in Fig. \ref{fig:runtimescomparison} that for these galaxies the run-times of \clustar\ are at least $3$ orders of magnitude faster than those of \hoptics\ and this disparity grows for larger data sets. \clustar\ follows an $\mathcal{O}(n\mathrm{log}(n))$ time complexity while \hoptics\ {\it appears} to follow a $\mathcal{O}(n^{2.17})$ time complexity. This is not the true time complexity of \hoptics\ and is partially an artefact of the changing clustering structuring within the galaxies -- a feature that affects both algorithms albeit affecting \clustar\ to a much lesser degree.

Given two data sets with an equal number of points, the run-time of \hoptics\ will be larger for the data set that has a larger range of densities within the virial radius i.e. is more cusp-like rather than core-like. This is because there will be a larger fraction of points whose mutual distance is less than or equal to $\epsilon$. Ordinarily, the best-case run-time of \hoptics\ is $\mathcal{O}(n\mathrm{log}(n))$, however as the larger synthetic galaxies we use in this paper are more cusp-like, we find that the time complexity of \hoptics\ approaches $\mathcal{O}(n^2)$ with increasing $n$. So while it is likely that the run-time of \hoptics\ applied to each galaxy individually is sub-quadratic, the overall trend of increasing cupsyness gives rise to a super-quadratic time complexity of $\mathcal{O}(n^{2.17})$.

As \clustar\ does not perform a radial search it does not suffer from this same drawback and moreover, the clustering structure does not strongly affect the run-time of the $k_\mathrm{den}$ nearest neighbour search and density computation either. In reality the run-time of \clustar\ in Fig. \ref{fig:searchsubstructure} is dominated by the $k_\mathrm{den}$ nearest neighbour search -- taking up a constant $\sim89$\% of the run-time for this number of dimensions and this value of $k_\mathrm{den}$. In this implementation, \clustar\ uses \code{SciPy}'s cKDTree \citep{Bentley1975, Maneewongvatana1999, SciPy} which has an expected build and search run-time of $\mathcal{O}(n\mathrm{log}(n))$. Such a run-time complexity is theoretically the fastest that can be achieved for this problem and hence \clustar's overall run-time only stands to improve by some constant factor. Faster run-times can be achieved this way by running the $k_\mathrm{den}$ nearest neighbour search in parallel, by using a faster implementation of the kdtree, or by using an approximate nearest neighbour algorithm rather than an exact one.

In the event that these run-time improvements are used and the dimensionality of the data set is sufficiently low, it is possible that the run-time of $k_\mathrm{den}$ nearest neighbour search becomes small enough that it is no longer the most time consuming component of the algorithm. The next most time consuming component of the \clustar\ algorithm is the aggregation and rejection process in Fig. \ref{fig:aggregate} which takes up a constant $\sim10$\% of the total run-time for each of the runs in Fig. \ref{fig:runtimescomparison}. This run-time is predominantly affected by the merging of potential clusters. In the best-case run-time scenario, there will be no mergers and the time complexity for this part of the algorithm would be $\mathcal{O}(n)$. While technically it is possible for a data set to be structured in such a way that gives rise to no mergers during the aggregation process, for astrophysical data, it would be highly atypical. Most if not all points in the data set will be densely connected together during this process. As the mergers are discovered they are combined in a tree-like structure. By merging the densely connected lists in a compact binary-tree structure the aggregation process will be performed in an $\mathcal{O}(n\mathrm{log}(n))$ time. This is both the expected and worst case time complexity. The time complexity of the aggregation process is not only affected by the way in which the densely connected lists are merged but also by how many times this occurs. The values $k_\mathrm{den}$ and $k_\mathrm{link}$ both affect this number and will increase the run-time by some constant factor the smaller that they are.

The remaining $\lesssim1$\% of \clustar's run-time can be attributed to the transformation of the data and general data manipulation -- the former of which has an $\mathcal{O}(nd^2 + d^3)$ time complexity. Overall, the these constituents make the run-time complexity of \clustar\ $\mathcal{O}(n\mathrm{log}(n))$. In Fig. \ref{fig:runtimescomparison} we do not show the run-time of \clustar\ when $adaptive = 0$ since this is only slightly different ($\lesssim1$\% different as above) from when $adaptive = 1$. We see that when $adaptive = 2$ the run-time increases by some factor per data set. In practice, this factor that is less than the number of levels found within the clustering hierarchy. When applied to the galaxies we analyse in this paper, the depth of the hierarchy is typically capped at $3$ or $4$ levels. \clustar's fast run-time makes it ideal for application to large data sets such as the Gaia DR3 catalogue.

\section{Parameter Influence and Optimisation} \label{sec:optimisation}
The \clustar\ algorithm has $7$ parameters. Parameters $l_x$, $adaptive$, and $k_\mathrm{den}$ may be chosen by the user and are responsible for the spatial linking length that finds field haloes, the adaptivity of the distance metric, and the number of neighbours used to find the density of the points respectively. A typical choice for $l_x$ is $0.2$ times the inter-particle spacing and then whether $adaptive$ is set to $0$, $1$, or $2$ is decided in consideration of the run-time vs clustering power trade-off as well as whether the user wishes to apply a PCA transform to the input data.

The parameter $k_\mathrm{den}$ has a more loosely constrained range -- if it is too small then the density fluctuations between neighbouring points can be large and noisy; if it too large then significant differences in local density may be smoothed out enough that some structures may not be detected by \clustar. Typically, similar parameters to $k_\mathrm{den}$ in other clustering algorithms are chosen with values ranging from $20$ to $500$. It is commonly accepted that $k_\mathrm{den}$ should increase when using data with a larger a feature space -- however this depends on the intended resolution of the density estimate and, in a clustering context, the expected number of points within the true clusters present in the data. A more thorough investigation of the effects of the choice of $k_\mathrm{den}$ is carried out in Secs. \ref{subsec:rhothresholdvalues} -- \ref{subsec:frejectvalues}.

Although the remaining $4$ parameters in \clustar\ may be chosen by the user, these do have properties that allow for their optimal values to be determined. It should be mentioned that these parameters do have some co-dependence with regards to the state of the output, though this influence is small. Due to the fact that none of these parameters act within the same step of the algorithm, we are able to isolate their effects in order to find their optimal values. Furthermore and since the labels within the synthetic data provide a flat clustering without noise, we opt to assess the performance of \clustar\ -- and hence the corresponding optimal values of some of the cluster extraction parameters ($\rho_\mathrm{threshold}$ and $S_\mathrm{outlier}$) -- by determining how well the leaf clusters of \clustar\ are able to match a single label within the synthetic data sets.

While the affects of the cluster extraction parameters in \clustar\ are expected to be similar to those in \hoptics, the lists of points that they act upon are generally slightly different. The near-optimal values of \hoptics\ parameters were determined by maximising the recovery and purity fractions of a set of hand-crafted 3D Gaussian distributions in both Cartesian and spherical coordinate systems. While this technique can be used to mimic returning the optimal set of astrophysical clusters, it of course is not equivalent to doing so. Furthermore, this investigation did not analyse the effects of higher dimensions and/or other subspaces on these parameters. We now investigate the effects of these parameters and deduce their preferred values.

\subsection{\texorpdfstring{Values for $\boldsymbol{k_\mathrm{link}}$}{Values for klink}} \label{subsec:klinkvalues}
The parameter $k_\mathrm{link}$ is the number of neighbours used to aggregate the points together in order of decreasing density so that the substructure may be found. The consequence of decreasing $k_\mathrm{link}$ is that the clusters will be more complete up until their true boundary density. Irrespective of this however, a practical lower limit exists. During the aggregation process, multiple potential clusters will be connected together into a new potential parent cluster whenever the current point being processed has multiple of its $k_\mathrm{link}$ nearest neighbours in different potential clusters. This has the consequence that at most $k_\mathrm{link} - 1$ potential clusters can be connected together during this step (since the point being processed is one of its own neighbours and does not yet belong to any potential cluster). The number of potential clusters that can be joined depends on the specific arrangement of points -- although with a sufficiently large value of $k_\mathrm{link}$, the number of these clusters that are observed to be joined together in these steps is most commonly $2$, occasionally $3$, and rarely $4$. The is also a subtle dependency on $k_\mathrm{den}$ as well, however the number of merging clusters only starts to increase for small values of this parameter i.e. $< 20$. Given that typically $k_\mathrm{den} \geq 20$, we suggest that for the choice of $k_\mathrm{link}$ not to affect this behaviour, its value needs to be at least as large as $5$ or $6$.

Furthermore, we must also consider that certain arrangements of points may give arise to the issue whereby the $k_\mathrm{link}$ nearest neighbours of each point is not an extensive enough neighbourhood for the process to be able to seamlessly aggregate most, if not all, points together. Densely connecting all points via their neighbourhood in a data set is not always feasible. In particular, if all points in the data set are contained within dense clusters that are each sparsely distributed with respect to each other then connecting all points may require a value of $k_\mathrm{link}$ that approaches the size of the data set itself.

\begin{figure}
    \centering
    \includegraphics[trim={6mm 15mm 13mm 29mm}, clip, width=\columnwidth]{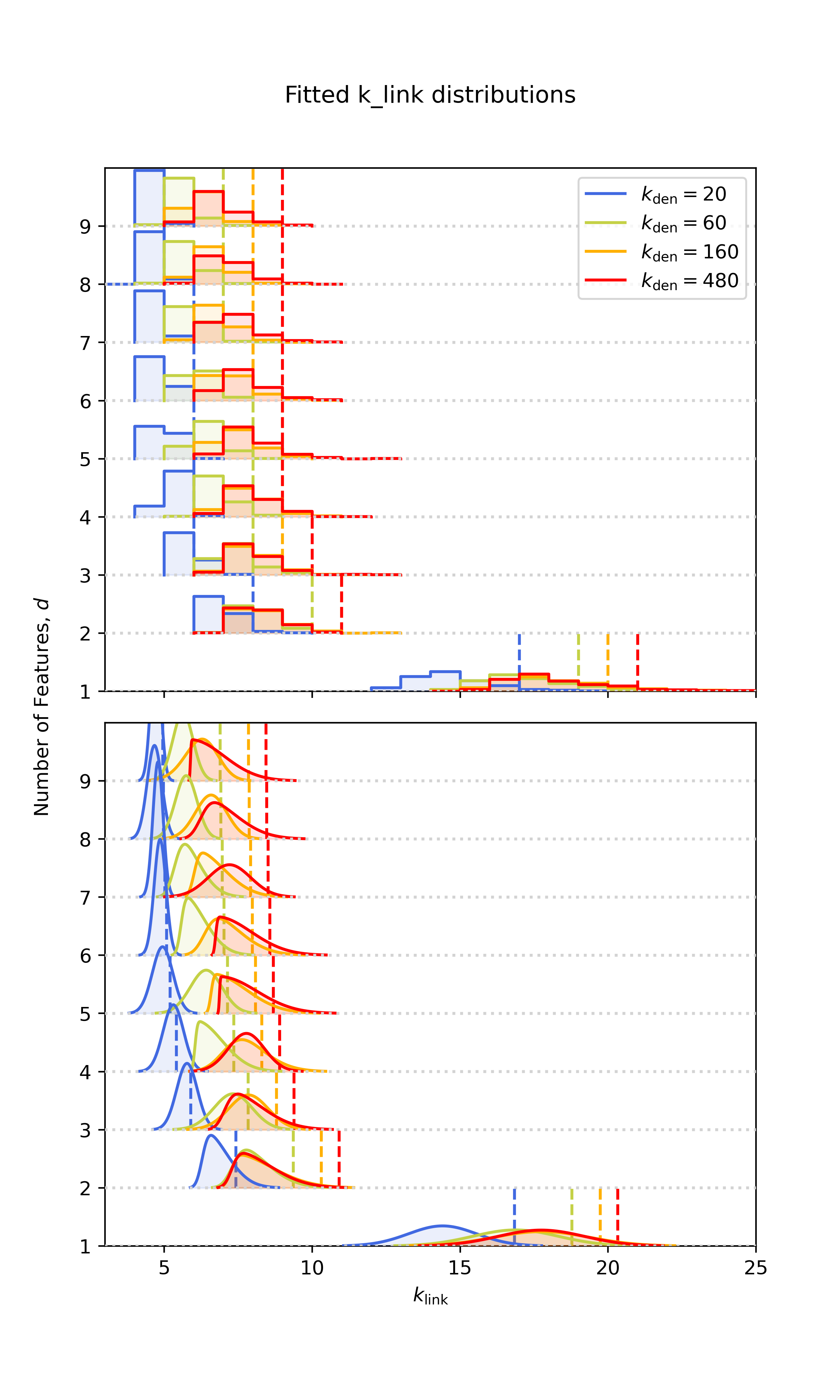}
    \vspace{-0.5cm}
    \caption{The empirical distributional relationships between the minimum $k_\mathrm{link}$ required to connect all $10^5$ points of a $d$-dimensional uniform distribution contained to the unit hypercube given that $k_\mathrm{den}$ nearest neighbours are used to estimate the local density. The value of $d$ is varied from $1$ to $9$ and the value of $k_\mathrm{den}$ is varied within the set $\{20,\ 30,\ 40,\ 60,\ 80,\ 120,\ 160,\ 240,\ 360,\ 480\}$ -- however only specific values of the latter are shown. The histograms in the top panel are the binned minimum $k_\mathrm{link}$ values and are created from $300$ re-samplings of the aforementioned data distribution for each $d$ and $k_\mathrm{den}$ combination. The skew-Gaussian distributions in the bottom panel are the fitted continuous analogues of the histograms in the top panel. The dashed lines in the lower panel are the $95^\mathrm{th}$ percentile values of the fitted skew-Gaussian distributions and the dashed lines in the top panel are the discrete analogue of the latter.}
    \label{fig:klink_values}
\end{figure}

To manage this issue, we choose practical values for $k_\mathrm{link}$ by ensuring that \clustar\ is able to densely connect all points given that they are from a randomly sampled uniform distribution in $d$-dimensions when it is using a particular value for $k_\mathrm{den}$. We vary both $d$ consecutively from $1$ to $9$ and $k_\mathrm{den}$ within the set $\{20,\ 30,\ 40,\ 60,\ 80,\ 120,\ 160,\ 240,\ 360,\ 480\}$. For each $d$-$k_\mathrm{den}$ combination we create $300$ random samples of $10^5$ points from a $d$-dimensional uniform distribution within the unit hypercube. For each of these we find the smallest value of $k_\mathrm{link}$ that will densely connect all $10^5$ points together by the end of the aggregation process. The histograms of these smallest values of $k_\mathrm{link}$ can be seen in the top panel of Fig. \ref{fig:klink_values} for various combinations of $d$ and $k_\mathrm{den}$. Within these histograms a value of $k_\mathrm{link} = x$ will contribute to the height of the bar in the interval $(x - 1, x]$ \footnote{Constructing the histogram in this way helps to provide a continuous analogue of the probability distribution since in such an analogue the value of $k_\mathrm{link} = x$ will have been drawn from this interval.}.

We then fit a skew-normal such that the sum of squared differences between the probability within each of the histogram's intervals and the probability of the skew-Gaussian within those same intervals is minimised. The skew-normal distribution we fit has the standard form of
\begin{align}
    &f(k_\mathrm{link}) = \frac{2}{\omega}\phi\bigg(\frac{k_\mathrm{link} - \xi}{\omega}\bigg) \Phi\bigg(\alpha\frac{k_\mathrm{link} - \xi}{\omega}\bigg),\ \mathrm{where} \notag\\
    &\phi(x) = \frac{1}{\sqrt{2\pi}}e^{-\frac{x^2}{2}},\ \mathrm{and} \notag\\
    &\Phi(x) = \int_{-\infty}^x \phi(y)dy = \frac{1}{2}\bigg[1 + \mathrm{erf}\bigg(\frac{x}{\sqrt{2}}\bigg)\bigg].
\end{align}
Here, $\xi$, $\omega$, and $\alpha$ are the location, shape, and scale parameters respectively. The continuous distributions shown in the bottom panel of Fig. \ref{fig:klink_values} depicts the fitted skew-normal distributions for various combinations of $d$ and $k_\mathrm{den}$.

We now use the fitted skew-normal distributions to find the continuous analogue value for $k_\mathrm{link}$ that predicts that all $10^5$ points within each uniform distribution will be densely connected together $95$\% of the time. Once found, we then fit a function of the form $k_{\mathrm{link},\ .95} = \alpha d^\beta + \gamma{k_\mathrm{den}}^\delta + \epsilon$ by minimising the squared differences between these distributional $95^\mathrm{th}$ percentile values and the function's output -- finding that $\alpha \approx 12.0$, $\beta \approx -2.2$, $\gamma \approx -23.0$, $\delta \approx -0.6$, and $\epsilon \approx 9.0$. The predicted continuous analogue value of $k_{\mathrm{link},\ .95}$ are shown as dashed vertical lines in the bottom panel of Fig. \ref{fig:klink_values} for various combinations of $d$ and $k_\mathrm{den}$. We then convert these back to the original discrete regime by rounding up to the nearest integer. Likewise, the discrete values of $k_{\mathrm{link},\ .95}$ are shown as dashed vertical lines in the top panel of Fig. \ref{fig:klink_values} for various combinations of $d$ and $k_\mathrm{den}$.

The fitted values of $k_{\mathrm{link},\ .95}$ in the continuous analogue do not always correspond well to the true $95^\mathrm{th}$ percentile of the fitted skew-normal distributions, however we find that they do correspond well to the empirical $95^\mathrm{th}$ percentile of histograms produced over each combination of $d$ and $k_\mathrm{den}$. To be more certain that the automation of $k_\mathrm{link}$ will not artificially break densely connected regions we add $1$ to these discrete values of the $95^\mathrm{th}$ percentile and -- to keep the potential cluster merging behaviour unaffected as discussed above -- always ensure that it is larger than $7$. To be clear, unless the user specifies a value for $k_\mathrm{link}$ \clustar\ calculates it using automatically using
\begin{equation} \label{eq:finalklink}
    k_\mathrm{link} = \mathrm{max}\{\mathrm{ceil}(12.0d^{-2.2} - 23.0{k_\mathrm{den}}^{-0.6} + 10.0),\ 7\}.
\end{equation}

Constructing $k_\mathrm{link}$ this way standardises the degree with which it is possible to connect the points of a data set whilst also ensuring maximal cluster completeness and as such these calculated values serve as a practical lower limit for $k_\mathrm{link}$. The user may choose different values for $k_\mathrm{link}$, although there are some considerations. The value of $k_\mathrm{link}$ must always be smaller than or equal to that of $k_\mathrm{den}$ -- which applies to the user's choice of $k_\mathrm{den}$ just as it does to that of $k_\mathrm{link}$. In the event that $k_\mathrm{link}$ is chosen automatically by \clustar\ the above optimal values are effectively the lower limits of $k_\mathrm{den}$ given the number of dimensions $d$. The value of $k_\mathrm{link}$ could be chosen to be larger than the automated values for the purpose of decreasing the run-time, however this comes at the detriment of cluster completeness. Other than these considerations, the optimal values of $k_\mathrm{link}$ do not depend on the choice of $k_\mathrm{den}$ nor any other parameter from \clustar\ as they have been determined via the simple criterion that all points in a $d$-dimensional uniform distribution be densely connected.

\begin{figure*}
    \centering
    \includegraphics[trim={16mm 5mm 23mm 16mm}, clip, width=0.9\textwidth]{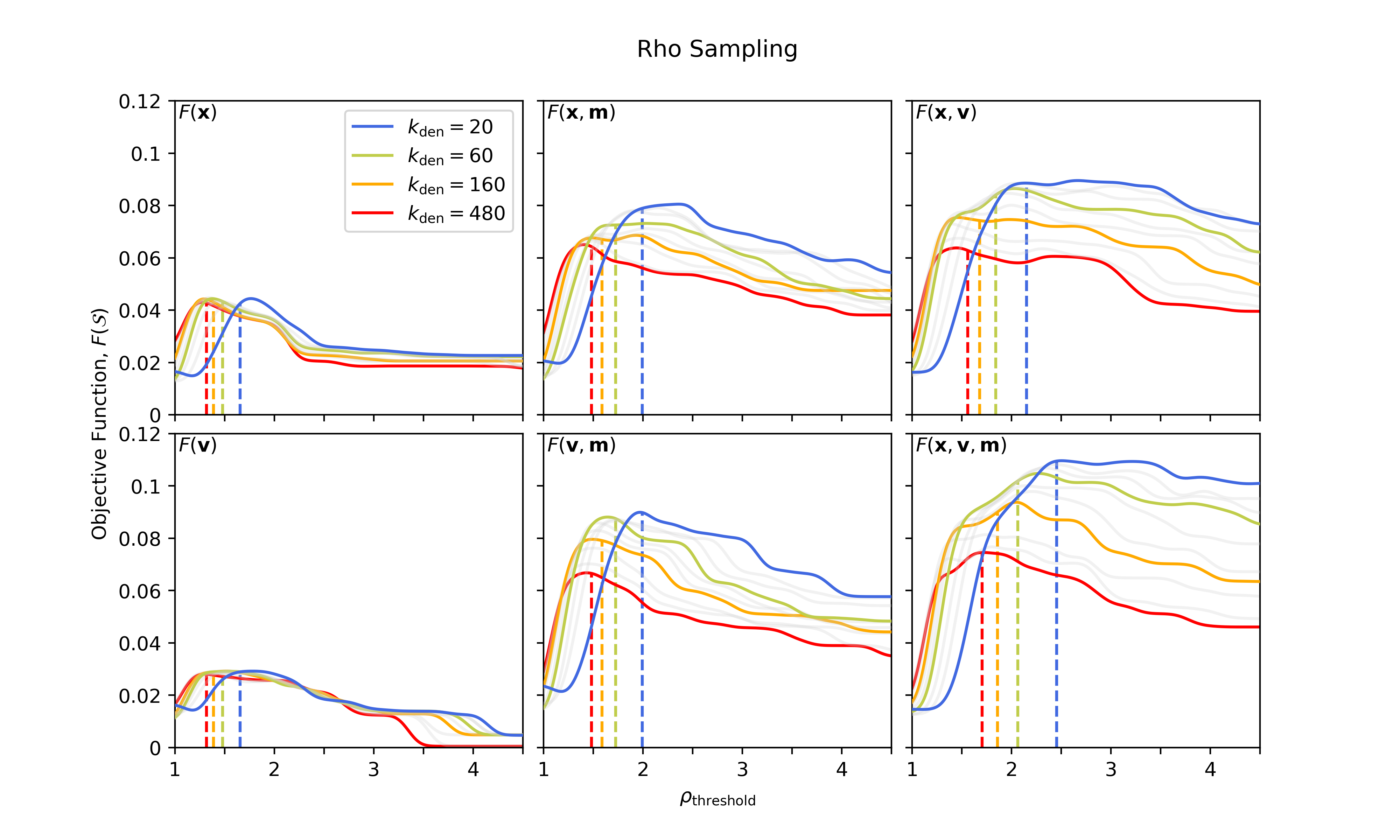}
    \vspace{-0.1cm}
    \caption{The clustering power of \clustar\ as it depends on the various feature subspaces, $k_\mathrm{den}$, and $\rho_\mathrm{threshold}$. The objective function used here is defined in Eq. \ref{eq:objectivefunction} however the plots show a Gaussian-smoothed version of this for simplicity. The dashed lines indicate the fitted values of $\rho_\mathrm{threshold}$ as a function the number of features, $d$, and the value of $k_\mathrm{den}$. These fitted values are calculated using Eq. \ref{eq:rho_fit}.}
    \label{fig:rho_sampling}
    \vspace{-0.5cm}
\end{figure*}

\subsection{\texorpdfstring{Values for $\boldsymbol{\rho_\mathrm{threshold}}$}{Values for rhothreshold}} \label{subsec:rhothresholdvalues}
The parameter $\rho_\mathrm{threshold}$ is used to provide the functionality of determining which groupings of points are {\it significantly} dense with respect to their surrounds. More specifically, it is responsible for ensuring that the median density of any cluster is at least $\rho_\mathrm{threshold}$ times the density of that cluster's surrounds -- refer to Eq. \ref{eq:rhocondition}.

For its use in \clustar, we take a closer look at the $\rho_\mathrm{threshold}$ parameter by optimising it on the synthetic galaxies described in Sec. \ref{sec:syntheticdata}. Since the labels within these data sets only provide a flat clustering of the synthetic galaxies without noise -- instead of a hierarchical clustering with noise as is the case with \clustar's output -- we only use the leaf clusters found by \clustar\ in order to determine its performance with varying $\rho_\mathrm{threshold}$. This approach then requires $f_\mathrm{reject}$ to be sufficiently large -- to ensure that the largest leaf cluster shares less than $f_\mathrm{reject}$ of it's points with the root cluster -- so that the leaf clusters will be unaffected by the choice of $f_\mathrm{reject}$. In fact, this poses a practical lower limit to $f_\mathrm{reject}$, however we revisit this with greater detail in Sec. \ref{subsec:frejectvalues}. Here, we set $f_\mathrm{reject} = 1$. Since the effect of $S_\mathrm{outlier}$ on the final clusters is only minimal, we simply set $S_\mathrm{outlier} = \infty$ to remove its influence -- we take a closer look at $S_\mathrm{outlier}$ in Sec. \ref{subsec:soutliervalues}. The $k_\mathrm{link}$ parameter is automatically chosen and accordingly to the scheme outlined in Sec. \ref{subsec:klinkvalues}. For simplicity, we also optimise these parameters using $adaptive = 1$.

By setting the above parameters in this way, we are able to find the optimal values of $\rho_\mathrm{threshold}$ as they depend on $k_\mathrm{den}$ and the number of features, $d$. We vary $\rho_\mathrm{threshold}$ from $1$ to $4.5$ in intervals of $0.1$ and $k_\mathrm{den}$ again from the set of $\{20,\ 30,\ 40,\ 60,\ 80,\ 120,\ 160,\ 240,\ 360,\ 480\}$. For each combination of $\rho_\mathrm{threshold}$ and $k_\mathrm{den}$, we run \clustar\ over each of the various feature subspace combinations out of the spatial ($\bf{x}$), kinematic ($\bf{v}$), and chemical ($\bf{m}$) subspaces\footnote{Note that we do not analyse the clusterings over the $\bf{m}$-subspace alone as typically each data point shares the same values of both [Fe/H] and [$\alpha$/Fe] with a number of other data points. Producing clusterings over the $\bf{m}$-subspace thereby artificially predicts many thousands of spurious clusters that only exist in this context because each point they contain is effectively a duplicate -- this dramatically increases the estimation of their local density and the density contrast between {\it nearby} neighbourhoods.

It is likely that this will also affect the clusterings in higher dimensional feature spaces that contain the $\bf{m}$-subspace to some extent. However, since the number of points per value of the ([Fe/H], [$\alpha$/Fe]) tuple is small (compared to the total number of points) in each synthetic galaxy, the global Mahalanobis metric used in \clustar\ here is effectively blind to the fact the data sets in higher dimensions are made up many hyper-planes separated along the [Fe/H] and [$\alpha$/Fe] directions. Specifically, the blindness that \clustar\ has to this bias in higher dimensional settings is a result of the PCA transformation where the resultant inter-point spacing within each local hyper-plane is on the order of the inter-plane spacing between each hyper-plane.}. Then for every run, we compare the leaf clusters found by \clustar\ to the ground truth labels from the relevant synthetic galaxy. We use the mutual-information-based objective function, $F(\mathcal{S})$, introduced in Sec. \ref{subsubsec:clusteringpower} and Eq. \ref{eq:objectivefunction} to assess the {\it clustering power} of \clustar\ for each $\rho_\mathrm{threshold}$ -- $k_\mathrm{den}$ -- subspace combination.

Fig. \ref{fig:rho_sampling} contains a panel for each feature subspace combination whereby $F(\mathcal{S})$ is plotted against the values of $\rho_\mathrm{threshold}$ for each value of $k_\mathrm{den}$ (with specific values shown in colour). As is mentioned in Sec. \ref{sec:syntheticdata}, the spatial and kinematic features (${\bf x}$ and ${\bf v}$) are each $d = 3$ subspaces and the chemical features (${\bf m}$) is a $d = 2$ subspace. From these plots we can clearly see a inter-dependency between $\rho_\mathrm{threshold}$, $k_\mathrm{den}$, and the number of features, $d$. As such, we wish to find a function that describes the optimal values for $\rho_\mathrm{threshold}$ such that this function is monotonically increasing with $d$ and monotonically decreasing with $k_\mathrm{den}$. We also need for $\rho_\mathrm{threshold}$ to always be at least $1$ as there is no reason for choosing otherwise.

In light of these factors and $\rho_\mathrm{threshold}$'s coupled dependency on $k_\mathrm{den}$ and $d$ (unlike the inter-dependency between these variables and $k_\mathrm{link}$ investigated in Sec. \ref{subsec:klinkvalues}), we choose to fit a function of the form $\rho_\mathrm{threshold}(k_\mathrm{den}, d) = 1 + \alpha f(d)/g(k_\mathrm{den})$ where $f$ and $g$ are monotonically increasing and strictly non-zero for positive inputs. We consider each power-law/logarithmic form combination for $f$ and $g$, but ultimately find that setting $f(d) = d^\beta$ and $g(k_\mathrm{den}) = \mathrm{ln}(k_\mathrm{den})$ gives the best overall fit. In fitting, we maximise the function's output on all Gaussian-smoothed objective function curves (for all values of $k_\mathrm{den}$ not just those that are coloured) shown in Fig. \ref{fig:rho_sampling} simultaneously with an equal weighting placed on each. Following this, we find that the optimal values of $\rho_\mathrm{threshold}$ are well-described by the function
\begin{equation} \label{eq:rho_fit}
    \rho_\mathrm{threshold}(k_\mathrm{den}, d) = 1 + 0.81\frac{d^{0.81}}{\mathrm{ln}(k_\mathrm{den})}.
\end{equation}

Some fitted values produced by Eq. \ref{eq:rho_fit} are shown in Fig. \ref{fig:rho_sampling} as vertical dashed lines with colours that correspond to the appropriate value of $k_\mathrm{den}$. While these fitted values are not always at the global maxima of the objective function curves, the function does not sacrifice much relevant information for the added convenience of having this parameter be chosen automatically from the user's choice of $k_\mathrm{den}$ and the number of features, $d$, in the data set that the user wishes to produce a clustering from.

In addition to being able to justify a set of optimal choices for $\rho_\mathrm{threshold}$, we can also see from Fig. \ref{fig:rho_sampling} the difference in information content as it depends on the specific feature subspace combination present within the input data. Here we can easily see that including extra informative dimensions will increase clustering power. We can also see that smaller $k_\mathrm{den}$ values produce better clusterings in an astrophysical context since they increase the resolution with which \clustar\ can resolve structure. This is particularly prevalent in the larger feature spaces where fine-structure can be separated from background noise more easily -- however choosing small values for $k_\mathrm{den}$ comes with diminishing returns as this also increases the effects of Poisson noise on the density estimation which can artificially introduce spurious structures. We take a closer look at the robustness of the \clustar\ output by comparing recovery and purity statistics of specific clusters across these same feature subspace combinations in Sec. \ref{sec:infocontent}.

\begin{figure*}
    \centering
    \includegraphics[trim={16mm 5mm 23mm 16mm}, clip, width=0.9\textwidth]{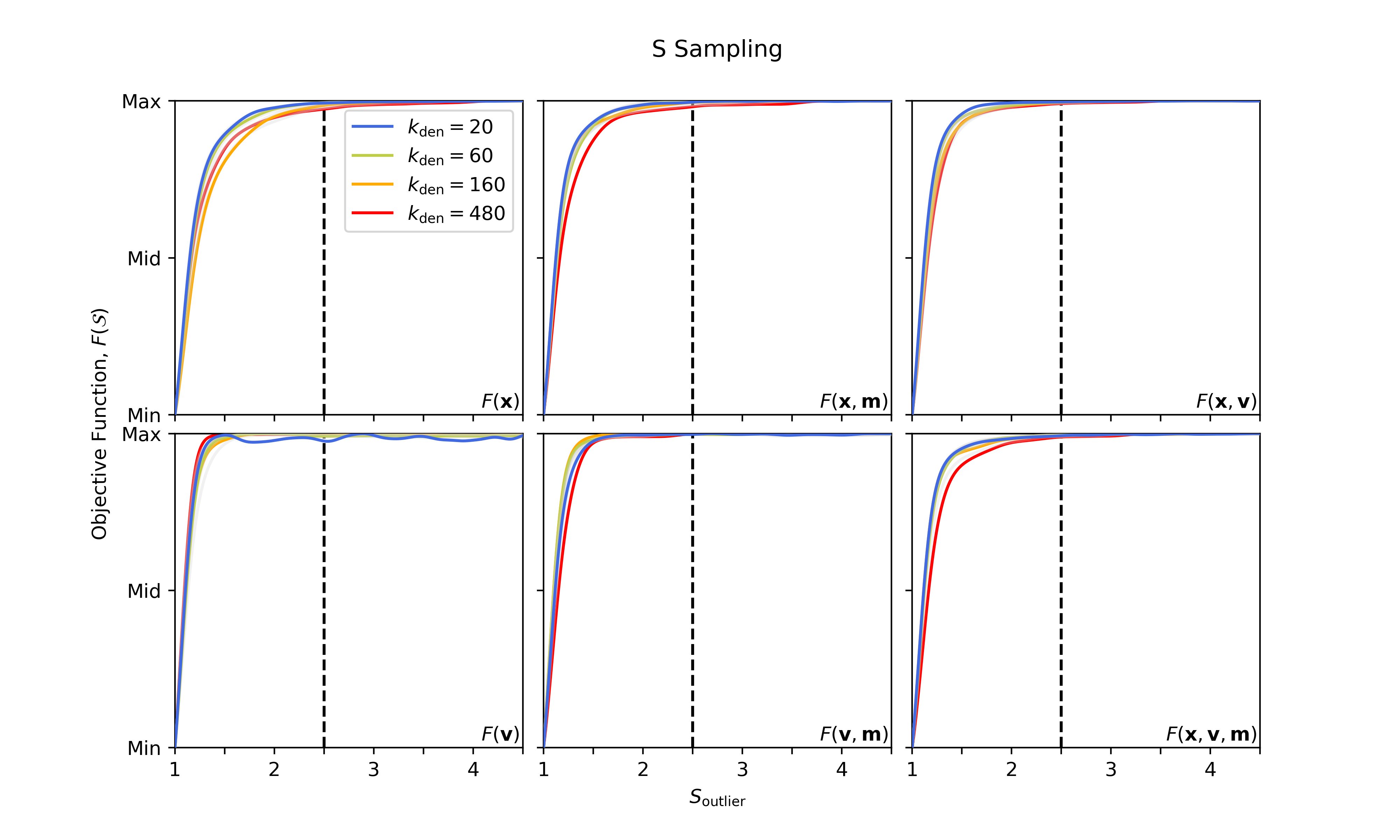}
    \vspace{-0.1cm}
    \caption{The clustering power of \clustar\ as it depends on the various feature subspaces, $k_\mathrm{den}$, and $S_\mathrm{outlier}$. The objective function used here is defined in Eq. \ref{eq:objectivefunction} and is re-normalised between each curve's respective minimum and maximum over this range for $S_\mathrm{outlier}$ to better depict the global influence of this parameter. As with Fig. \ref{fig:rho_sampling}, these plots show a Gaussian-smoothed version of the objective function. We see that for small $S_\mathrm{outlier}$ values the clustering power decreases dramatically and for large values the clustering power asymptotically approaches a maximum (values of which are those shown in Fig. \ref{fig:rho_sampling} at the relevant $\rho_\mathrm{threshold}$ values for each $k_\mathrm{den}$ and number of features, $d$). The dashed lines indicate the value of $S_\mathrm{outlier} = 2.5$ which we set as the standard choice for \clustar. This value is both small enough to remove distinct outliers from clusters while being large enough to not have the resultant clustering breakdown. Ultimately however, any value above about $S_\mathrm{outlier} = 1.5\sim2$ will produce good clustering results -- adjusting this further is dependent on how strict the user wishes to be on the removal of potential outliers from clusters.}
    \label{fig:s_sampling}
    \vspace{-0.5cm}
\end{figure*}

\subsection{\texorpdfstring{Values for $\boldsymbol{S_\mathrm{outlier}}$}{Values for Soutlier}} \label{subsec:soutliervalues}
The parameter $S_\mathrm{outlier}$ is used to categorise the points of the input data as either local inliers or local outliers depending on whether their local-outlier-factor (or rather its density kernel analogue in Eq. \ref{eq:lof}) is less than or greater than $S_\mathrm{outlier}$. Following the aggregation process described in Sec. \ref{subsec:aggregation}, a cut-off density is then defined for each cluster according to Eq. \ref{eq:rhocut}. This classification is then used to remove all points from each cluster whose local density is less than the cut-off density, $\rho_\mathrm{cut}$, of that cluster -- i.e. only local outliers whose density is less than that of all local inliers are removed.

To better understand the effect $S_\mathrm{outlier}$ has on the final clustering produced and gauge an optimal value for it from the synthetic galaxies discussed in Sec. \ref{sec:syntheticdata}, we compute the objective function, $F(\mathcal{S})$, in Eq. \ref{eq:objectivefunction} from the \clustar\ output for various combinations of $S_\mathrm{outlier}$, $k_\mathrm{den}$, and feature subspace in largely the same way as in Sec. \ref{subsec:rhothresholdvalues}. We vary $S_\mathrm{outlier}$ from $1$ to $4.5$ in intervals of $0.1$. We similarly vary $k_\mathrm{den}$ from the set of $\{20,\ 30,\ 40,\ 60,\ 80,\ 120,\ 160,\ 240,\ 360,\ 480\}$ and the feature subspace combinations of the spatial ({\bf x}), kinematic ({\bf v}), and chemical ({\bf m}) subspaces. Given these combinations of $k_\mathrm{den}$ and the number of features, $d$, we allow for $k_\mathrm{link}$ and $\rho_\mathrm{threshold}$ to be automatically chosen according to Eqs. \ref{eq:finalklink} and \ref{eq:rho_fit} respectively. In our investigation of $S_\mathrm{outlier}$ we again set $f_\mathrm{reject} = 1$ and $adaptive = 1$.

The relationship between the clustering power of \clustar\ and $S_\mathrm{outlier}$ is shown in Fig. \ref{fig:s_sampling} for each combination of $k_\mathrm{den}$ and $d$. More simplistic than that between the clustering power and $\rho_\mathrm{threshold}$, the relationship here is very similar between these combinations and suggests that so long as $S_\mathrm{outlier}$ is {\it large enough} the \clustar\ output will be robust\footnote{The $F({\bf v})$ panel of Fig. \ref{fig:s_sampling} appears to include a oscillation-like pattern for $S_\mathrm{outlier} \gtrsim 1.5$ and $k_\mathrm{den} = 20$. This is an artefact of the objective function minimum being closer to its maximum for this combination. As such, the scale of the noise appears much larger and noticeable compared to the other $k_\mathrm{den}$ and $d$ combinations. The exact reasoning for this noise is that for a given change in $S_\mathrm{outlier}$ the quality of some clusters will increase while others may decrease. These competing effects take place for every curve in Fig. \ref{fig:s_sampling} but are simply more noticeable on this curve due to the imposed scale between the minimum and maximum.}. For values larger than about $1.5$ or $2$, the choice of $S_\mathrm{outlier}$ has a minimal effect and will only serve to remove a select few or perhaps none of the points from each cluster. Of course changing $S_\mathrm{outlier}$ from $3$ and $4$ will still have an effect on which points are apart of the final clusters, however this effect is minimal and cannot be robustly defined beyond the resultant clustering include more potential outliers per cluster than beforehand. In light of this ambiguity, we choose to use $S_\mathrm{outlier} = 2.5$ as the standard value within \clustar\ as this will provide strong clustering results with a moderate level of outlier detection.

\subsection{\texorpdfstring{Values for $\boldsymbol{f_\mathrm{reject}}$}{Values for freject}} \label{subsec:frejectvalues}
The parameter $f_\mathrm{reject}$ defines the maximum fraction of points (relative to the parent cluster) that can be shared between any of the parent-child cluster pairs that exist immediately following the aggregation process. A cluster is rejected from a parent-child pair if the child shares more than $f_\mathrm{reject}$ of the parent's points -- according to Eq. \ref{eq:fcondition} -- i.e. \clustar\ removes the smallest cluster of the pair that has child clusters of its own, unless it would be a root-level cluster in which case the child from the pair is removed. These rules mean that $f_\mathrm{reject}$ is implicitly responsible for determining the size and shape of the hierarchy between the terminating clusters of the hierarchy. Refer to Sec. \ref{subsec:rejection} for more details on the rules surrounding the $f_\mathrm{reject}$ parameter.

Choosing extreme $f_\mathrm{reject}$ values of $0$ and $1$ would force \clustar\ to return $(\leq$\footnote{The $\leq$ sign here indicates that the number of leaf clusters found in the $f_\mathrm{reject} = 0$ case may be less than or equal to the number of leaf clusters found in the $f_\mathrm{reject} = 1$. The inequality will exist if there is one or more leaf clusters whose parent is the root-level cluster.}$)l$ and $2(l - 1)$ clusters per root-level cluster respectively -- where $l$ is the number of leaf clusters within that root-level cluster. In the same scenarios, the hierarchy returned by \clustar\ would have depths of $1$ and somewhere in the range of [$\mathrm{log}_2(l - 1)$, $l - 1$] per root-level cluster respectively. To use the standard terminology, these values of $f_\mathrm{reject}$ create hierarchies that form perfect $(\leq)l$-ary trees and full binary trees respectively -- with the exact shape of the latter dependent upon $l$'s representation in base-$2$ and the clustering structure within the data set. Varying $f_\mathrm{reject}$ between these values transforms the shape of the hierarchy in a way is difficult to predict.

We should expect that a reasonable lower limit for $f_\mathrm{reject}$ is $0.5$, since anything lower would suggest that equally sized groups can not be merged into a parent cluster. However, further constraining this within this paper would be a difficult task since the satellite labels within the synthetic data from \code{Galaxia} (Sec. \ref{sec:syntheticdata}) provide a flat clustering without noise of each galaxy and this information alone does not indicate the optimal hierarchy shape. To properly determine the optimal $f_\mathrm{reject}$ from this data, we would need to assess the physics of the galaxies and their constituents -- to determine how bound each satellite is to each other, and prepare an objective function that can be used to infer the most physical sense of what {\it a satellite within a satellite} should look like in an astrophysical context. Specifically, this would require analysis of the mass, orbits of the points, backwards integration, and analysis of the star formation histories within satellites and how this relates to each other satellite within a galaxy. Additionally, we can not even be certain from the outset that we would see a single most-optimal value for $f_\mathrm{reject}$ that is able to consistently provide such a hierarchy because the rules that govern this may not even be suitable to do so.

In light of these issues -- and since we expect the hierarchy produced by \clustar\ to be similar to that of \hoptics -- we choose to follow the verdicts with respect to the parameter $f_\mathrm{reject}$ in \citet{Oliver2020}. Considering the results from figure 6 therein, a value of $f_\mathrm{reject} = 0.9$ is near-optimal and ensures good quality clustering results (high Jaccard index as well as high recovery and purity fractions) of the hierarchical combinations of hand-crafted leaf clusters. From a hierarchy interpret-ability stand point, running \clustar\ with this value of $f_\mathrm{reject}$ over the synthetic galaxies discussed in Sec. \ref{sec:syntheticdata} creates consistently shaped hierarchies with levels $L = 1$, $2$, and $3+$ containing $\sim 70$\%, $\sim 20$\%, and $\sim10$\% of all galactic substructure clusters ($L > 0$). Furthermore, these percentages do not seem to change more than $\sim 15$\% regardless of the value of $k_\mathrm{den}$, the feature space of the input data, nor the choice synthetic galaxy -- implying that we should expect the resultant hierarchy shape to remain simple to interpret and predictable when \clustar\ is applied to astrophysical data sets.



\section{Information Content of Different Clustering Scenarios} \label{sec:infocontent}
In the preceding sections we have presented the \clustar\ algorithm (Sec. \ref{sec:clustar}), showing that it produces comparable clusters to \hoptics\ in a far more computationally efficient manner (Sec. \ref{sec:comparison}), and optimised its parameters for generalised application to n-dimensional data sets (Sec. \ref{sec:optimisation}). In this section we will further demonstrate the clustering power of \clustar\ and quantify the difference between the clusters that result when \clustar\ is applied to different galactic environments (Sec. \ref{subsec:galacticenvironments}), to different feature spaces (Sec. \ref{subsec:featurespace}), and using metric adaptivity settings (Sec. \ref{subsec:metricadaptivity}).

\begin{table}
\centering
\caption{A summary of the minima, medians, means, and maxima of the normalised mutual information values ($F_g(\mathcal{S})$) for the various feature spaces ($\mathcal{S}$) and $adaptive$ settings shown within Fig. \ref{fig:finalmutualinfo}. The mean values are equivalent to the objective function defined $F(\mathcal{S})$ in Eq. \ref{eq:objectivefunction}. For completeness, we also include these statistics for the clusterings produced with \hoptics\ using the $({\bf x})$ feature space in Sec. \ref{sec:comparison}.}
\begin{tabular}{|c|c|c|c|c|c|}
\hline
$\mathcal{S}$ & $adaptive$ & Min & Median & Mean & Max \\
\hline
\multirow{4}{*}{$({\bf x})$} & \hoptics & 0 & 0.027 & 0.043 & 0.198 \\
                             & $0$ & 0.001 & 0.028 & 0.043 & 0.197 \\
                             & $1$ & 0.006 & 0.027 & 0.043 & 0.193 \\
                             & $2$ & 0     & 0.023 & 0.039 & 0.193 \\
\hline
\multirow{3}{*}{$({\bf v})$} & $0$ & 0 & 0.009 & 0.029 & 0.261 \\
                             & $1$ & 0 & 0.008 & 0.029 & 0.266 \\
                             & $2$ & 0 & 0.006 & 0.028 & 0.265 \\
\hline
\multirow{2}{*}{$({\bf x}, {\bf m})$} & $1$ & 0.024 & 0.062 & 0.079 & 0.364 \\
                                      & $2$ & 0.017 & 0.066 & 0.075 & 0.341 \\
\hline
\multirow{2}{*}{$({\bf v}, {\bf m})$} & $1$ & 0 & 0.063 & 0.093 & 0.473 \\
                                      & $2$ & 0 & 0.070 & 0.102 & 0.469 \\
\hline
\multirow{2}{*}{$({\bf x}, {\bf v})$} & $1$ & 0 & 0.065 & 0.089 & 0.445 \\
                                      & $2$ & 0 & 0.067 & 0.092 & 0.443 \\
\hline
\multirow{2}{*}{$({\bf x}, {\bf v}, {\bf m})$} & $1$ & 0 & 0.077 & 0.110 & 0.485 \\
                                               & $2$ & 0 & 0.083 & 0.109 & 0.486 \\
\hline
\end{tabular}
\label{tab:finalobjfctn}
\end{table}
\begin{figure}
    \centering
    \includegraphics[trim={9mm 22mm 10mm 36mm}, clip, width=\columnwidth]{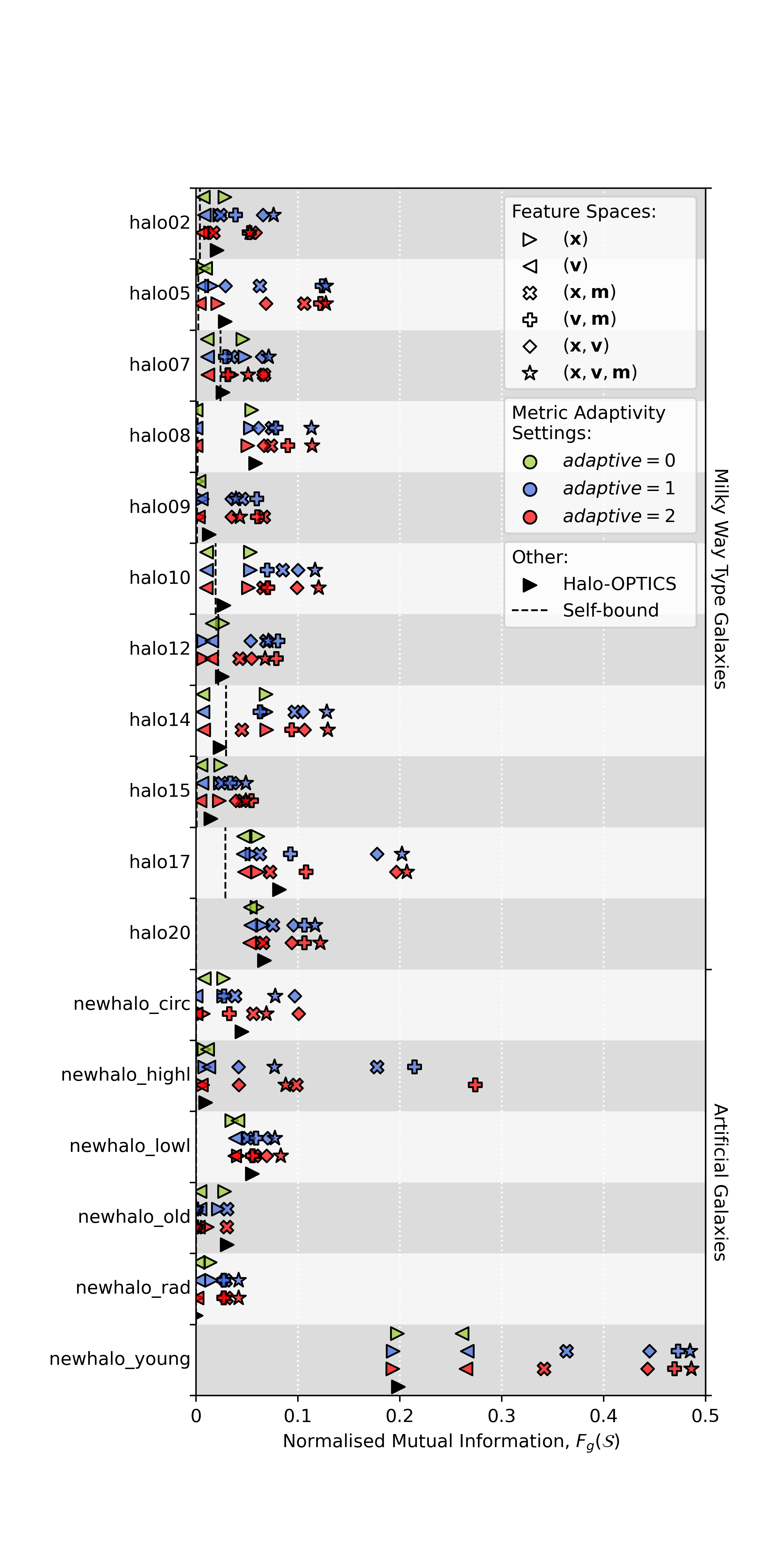}
    \vspace{-0.5cm}
    \caption{The normalised mutual information $F_g(\mathcal{S})$ from Eq. \ref{eq:objectivefunction} of various \clustar\ clustering scenarios involving different synthetic galaxies, feature spaces, and metric adaptivity settings. The $F_g(\mathcal{S})$ values of both the \hoptics\ clusterings from Sec. \ref{sec:comparison} and a hypothetical set of clusterings that perfectly classify ($J_\mathrm{max} = 1$) only the self-bound satellites within each MW type galaxy are also shown for reference. The vertical axes indicate the names of the synthetic galaxies as well as whether they are from the original MW type galaxies  of \citet{Bullock2005} or from the additional artificial galaxies of \citet{Johnston2008}. It is easy to see that the galactic environ, the number and type of informative features, and the metric adaptivity all affect the power of the resultant clustering.}
    \label{fig:finalmutualinfo}
\end{figure}

The clustering power of the now-optimised \clustar\ is summarised in Tab. \ref{tab:finalobjfctn} \& Fig. \ref{fig:finalmutualinfo}: Fig. \ref{fig:finalmutualinfo} illustrates the breakdown of the objective function from Eq. \ref{eq:objectivefunction} (i.e. $F_g(\mathcal{S})$) over each synthetic galaxy from Sec. \ref{sec:syntheticdata} -- the measure of normalised mutual information is each galaxy's contribution to the objective function prior to having averaged over them; Tab. \ref{tab:finalobjfctn} displays the minima, medians, means, and maxima of the distribution of $F_g(\mathcal{S})$ as it depends on the metric adaptivity setting and the feature space. With Fig. \ref{fig:finalmutualinfo} we can also see how the clustering power differs depending on the galactic environment as well. While these values may seem low, both Tab. \ref{tab:finalobjfctn} and Fig. \ref{fig:finalmutualinfo} show that \clustar\ has performed well given the circumstances of this data. As a reference point, Fig. \ref{fig:finalmutualinfo} also shows the clustering power of a hypothetical clustering that exclusively and perfectly classifies ($J_\mathrm{max} = 1$) the self-bound satellites from within each MW type galaxy. We see that in comparison, \clustar\ performs well and that the galaxies are composed of mostly disrupted tidal debris from previous mergers -- which is mostly impossible to aggregate together without joining multiple unbound groups and effectively reducing our understanding of the catalogue of true satellites. We see that even with \clustar\ only using the positions ($\mathcal{S} = ({\bf x})$) it will consistently return the self-bound satellites and some tidal debris to the effect of having an $F_g(\mathcal{S})$ value that is at least a factor of $\sim1$--$2$ times larger than for just classifying the self-bound satellites. This is unsurprising since \hoptics\ was shown to be able to do the same thing with only positional information \citep{Oliver2020}. As the feature space size increases \clustar's clustering power increases too such that when $\mathcal{S} = ({\bf x}, {\bf v}, {\bf m})$, $F_g(\mathcal{S})$ is typically at least a factor of $\sim3$--$6$ times larger than that for self-bound satellites only.

\begin{figure*}
    \centering
    \includegraphics[trim={24mm 16mm 23mm 16mm}, clip, width=0.9\textwidth]{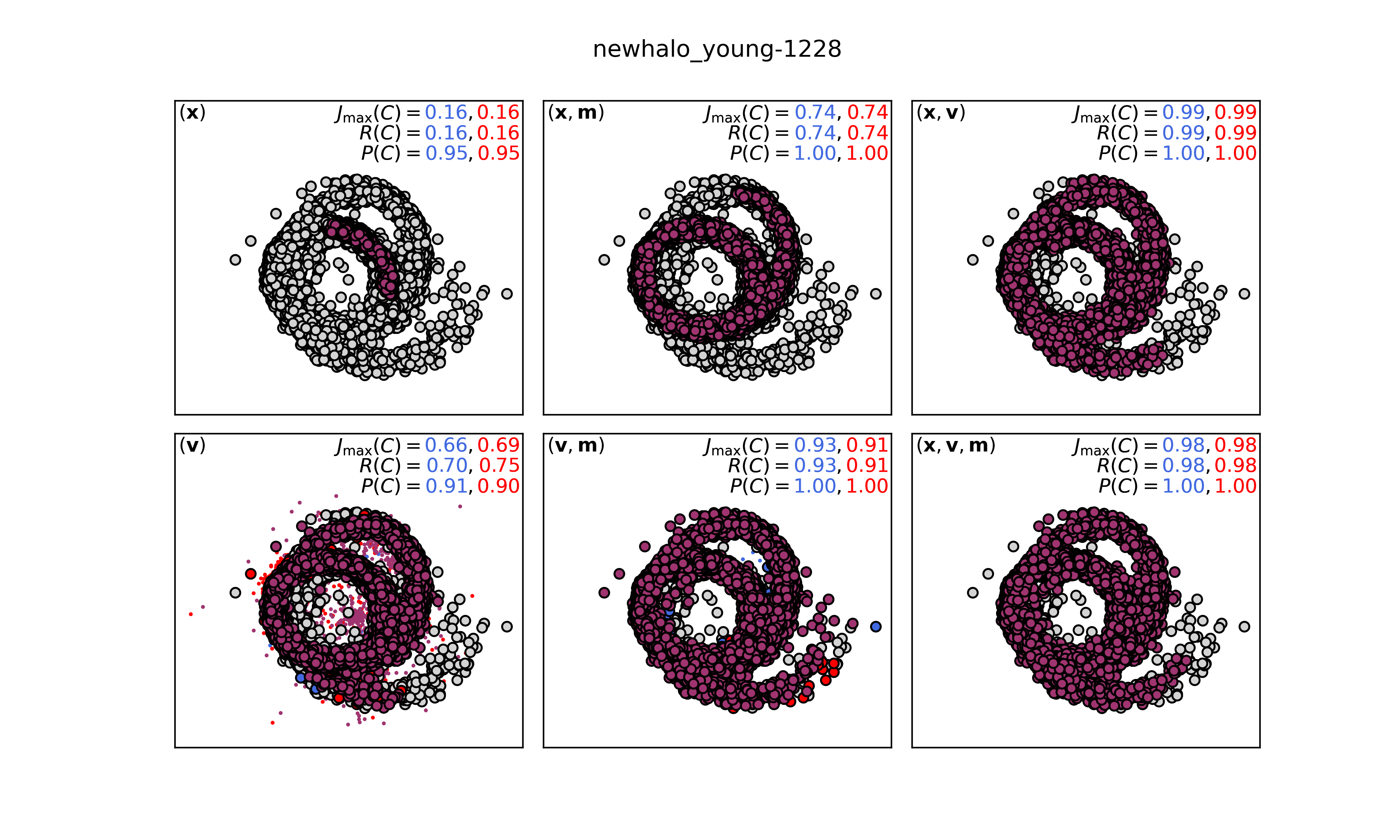}
    \vspace{-0.1cm}
    \caption{A satellite and the corresponding best-fitting clusters predicted by \clustar\ when using various feature spaces and metric adaptivity settings. The projection shown of the satellite in each panel is of the two spatial PCA components with the largest variances. The feature space used for the satellite's prediction is shown in the top left corner of each panel. The metric adaptivity setting is indicated by colour; where blue and red corresponds to an $adaptive$ value of $1$ and $2$ respectively -- the purple indicates their intersection. Large points belong to the satellite, small points are incorrectly associated to the satellite. In the top right corner of each panel are the maximum Jaccard index (Eq. \ref{eq:maxjaccardindex}), recovery, and purity (Eq. \ref{eq:recoveryandpurity}) of the best-fitting clusters from the two $adaptive$ values. These panels show the influence of the different feature spaces as well as that of the different metric adaptivity settings.}
    \label{fig:cluster_visual_1}
    \vspace{-0.5cm}
\end{figure*}

All clusters analysed in this section have been produced using $k_\mathrm{den} = 20$ -- as this consistently produced the best results in Sec. \ref{sec:optimisation}. All other parameters are chosen automatically according to the analysis in Sec. \ref{sec:optimisation} unless specified otherwise. We also highlight the impact of feature space and adaptivity setting using two satellites, shown in Fig. \ref{fig:cluster_visual_1} and \ref{fig:cluster_visual_2}.

\subsection{The Effects of Galactic Environment} \label{subsec:galacticenvironments}
Fig. \ref{fig:finalmutualinfo} shows that there are distinct differences in \clustar's clustering power that are dependent upon the {\it type} of galaxy that the algorithm is applied to. This is not unexpected, the constituent satellites of any particular galaxy will be more/less difficult to classify for any clustering algorithm if a large number of these satellites are disrupted/intact. In essence, if a large number of a satellite's points' $k_\mathrm{link}$ neighbourhoods contain points from multiple satellites then \clustar's ability to provide a well-matched cluster to that satellite diminishes as any matching predicted cluster will also contain contamination points from other satellites. Such galactic environments will occur if the constituent satellites have experienced strong tidal forces throughout their history or if they are sparsely distributed at the time of their infall such that they encompass a large volume of the feature space. It is for these reasons that we see the clustering power of \clustar\ to be less when applied to galaxies created by old satellites, satellites on radial trajectories, and low luminosity satellites.

Contrarily, if most $k_\mathrm{link}$ neighbourhoods contain only points from a single satellite then the aggregation process can likely link together many points from a single satellite in sequence before the group becomes overly contaminated by members of other satellites. Of course, \clustar's ability to create meaningful clusters from these neighbourhoods also depends on the density of the points and whether these neighbourhoods are isolated or connected to other like-neighbourhoods. Galactic environments that harbour such conditions will contain a larger proportion of intact satellites compared to those that do not. Accordingly, Fig. \ref{fig:finalmutualinfo} shows that galaxies made of young and (for some feature spaces) high luminosity satellites are more easily categorised into their respective satellites.

As the MW type galaxies are essentially a mix of disrupted and intact satellites we befittingly see a clustering power that lies somewhere in the middle of these extremes. Nevertheless, from these observations we can gain some understanding about what to expect if \clustar\ were to be applied to a data set of stars in the galactic bulge compared to if it were applied to the stars in the stellar halo.

\subsection{The Effects of Feature Space} \label{subsec:featurespace}
The effects of group-mixing in a data set are not only prevalent between different galactic environments, but are noticeable between different feature spaces as well. Generally speaking, Tab. \ref{tab:finalobjfctn} portrays that the \clustar\ clustering power is greater for larger feature spaces. One exception to this is that we find the $({\bf v}, {\bf m})$ feature space more informative than the $({\bf x}, {\bf v})$ feature space. Amongst equal sized feature spaces, we see that spatial coordinates are generally more informative than kinematic coordinates although the opposite is true when chemical abundances are added in amongst these feature spaces.

From Fig. \ref{fig:finalmutualinfo} however, we see that the above observations from Tab. \ref{tab:finalobjfctn} are not consistent across galaxies -- i.e. the information content of any particular galaxy clustering is coupled to both the galactic environment and the underlying feature space of the data. Out of the synthetic galaxies we investigate in this paper, the clustering scenario that gives the most information content for each galaxy is most commonly derived from the largest feature space, $({\bf x}, {\bf v}, {\bf m})$, however the title of {\it most informative feature space} is also sometimes held by the $({\bf x}, {\bf m})$, $({\bf v}, {\bf m})$, or $({\bf x}, {\bf v})$ feature spaces.

\begin{figure*}
    \centering
    \includegraphics[trim={24mm 16mm 23mm 16mm}, clip, width=0.9\textwidth]{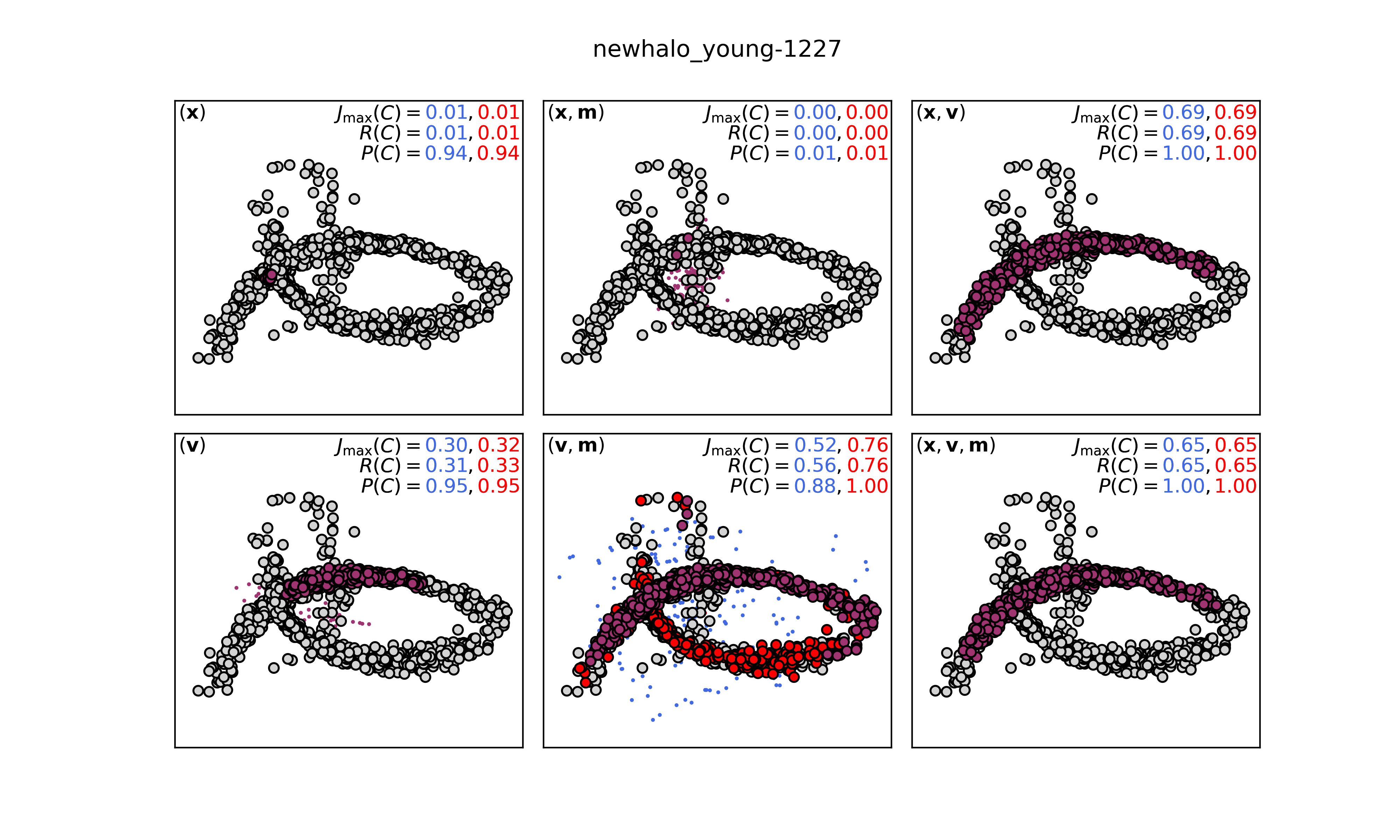}
    \vspace{-0.1cm}
    \caption{A satellite and the corresponding best-fitting clusters predicted by \clustar\ when using various feature spaces and metric adaptivity settings. This satellite is different to that shown in Fig. \ref{fig:cluster_visual_1}, however the description of each panel's information remains the same here. As with Fig. \ref{fig:cluster_visual_1}, these panels show the influence of the different feature spaces as well as that of the different metric adaptivity settings -- however the $({\bf v}, {\bf m})$ panel in the lower middle is a pronounced case of the clustering difference that can occur between the globally adaptive metric setting ($adaptive = 1$) and the locally adaptive metric setting ($adaptive = 2$).}
    \label{fig:cluster_visual_2}
    \vspace{-0.5cm}
\end{figure*}

Adding in extra features will often increase clustering power by creating some separation between multiple satellites within the extra volume that is provided by the additional dimensions. However, this will not always be the case as it is possible that by adding in extra {\it uninformative} features the points belonging to multiple satellites will be brought closer together (relative to other points) -- making it more difficult for \clustar\ to disentangle them from each other and their surrounds. These effects are why the clustering of the satellite depicted in Fig. \ref{fig:cluster_visual_1} is worse for the $({\bf x}, {\bf v}, {\bf m})$ feature space than it is for the $({\bf x}, {\bf v})$ feature space, even though the former feature space is more voluminous. The same effect can be also be seen in Fig. \ref{fig:cluster_visual_2}, not only between these feature spaces but between the $({\bf x})$ and $({\bf x}, {\bf m})$ feature spaces as well.

Irrespective of the above, Tab. \ref{tab:finalobjfctn} does confirm that this effect is outweighed by the opposite effect when including additional features to cluster over -- i.e. that including additional features brings together $k_\mathrm{link}$ neighbourhoods made of points from the same satellite while separating those that contain points from other satellites. Specifically, we can observe this occur in Fig. \ref{fig:cluster_visual_1} where the addition of the chemical features in the $({\bf x}, {\bf m})$ feature space has dramatically improved the best-fitting predicted cluster to the satellite when compared to that found using the spatial features alone. The same can be said about the $({\bf v}, {\bf m})$ and $({\bf v})$ feature spaces, and not only in Fig. \ref{fig:cluster_visual_1} but also in Fig. \ref{fig:cluster_visual_2}. Moreover, we see that the amalgamation of the spatial and kinematic features allows \clustar\ to produce far better clusterings of the respective satellites in both Figs. \ref{fig:cluster_visual_1} and \ref{fig:cluster_visual_2}.

One method of finding the most informative feature space is by comparing the Shannon entropy \citep{Shannon1948} of the available feature spaces. The smaller the Shannon entropy of a particular feature space the less uniform the distribution of points within it and hence more clustered -- this does not necessary guarantee that the feature space is appropriately clustered but it does hint to this possibility. Such a concept is already utilised within clustering algorithms such as \code{EnLink} \citep{Sharma2009} in order to create a meaningful locally adaptive metric and could be utilised when using \clustar\ to narrow the list of possible feature spaces before applying the algorithm.

\subsection{The Effects of Metric Adaptivity Settings} \label{subsec:metricadaptivity}
Another way to bring like-satellite points together while separating dislike-satellite points from each other is via the means of a adaptive metric. \clustar\ offers three metric adaptivity settings however only the $adaptive$ values of $1$ and $2$ should be considered as adaptive metrics since there is no transformation performed by \clustar\ when $adaptive = 0$. It is for this reason that the latter should only be used if the feature space consists solely of similar coordinates such as the $({\bf x})$ or $({\bf v})$ feature spaces. The $adaptive = 0$ setting can also be used on compound feature spaces so long as some meaningful transformation of the data has been performed. The settings of $adaptive = 1$ and $2$ are implementations of a globally adaptive metric and a locally adaptive metric respectively. 

In Tab. \ref{tab:finalobjfctn} we can see that running \clustar\ with the $adaptive = 1$ setting produces similar results to the $adaptive = 0$ in the feature spaces where this setting is applicable. This gives us confidence that the $adaptive = 1$ setting is appropriate for producing a globally adaptive metric for the larger compound feature spaces. The locally adaptive metric provides a similar clustering power overall to the globally adaptive metric. The medians and means of this measure appear to be higher for the locally adaptive metric when \clustar\ is applied to compound feature spaces, but it is difficult to say whether using it provides a better clustering in general. However, in Fig. \ref{fig:finalmutualinfo} we see that blanket statements about the use of the locally adaptive metric over the globally adaptive metric are not always true. With this it is clear that the optimal value of $adaptive$ has a subtle dependency on the galactic environment as well as the feature space.

Fig. \ref{fig:cluster_visual_2} depicts one such case where the use of the $adaptive = 2$ setting has caused \clustar\ to produce a distinctly better match to a satellite. We see here that by using the locally adaptive metric setting on the $({\bf v}, {\bf m})$ feature space, the best-fitting match to the satellite has both an improved recovery and purity. However it is possible that in other feature spaces, $({\bf x}, {\bf v})$ and $({\bf x}, {\bf v}, {\bf m})$ for example, the lower arm (on the satellite projection) has been found as a sibling cluster(s) to the best-fitting cluster(s) shown in those panels. In this sense, it is likely that the locally adaptive metric has simply created a means to condense the lower density particle bridge between these regions.

Intuitively, the $adaptive = 2$ setting is only preferable to the $adaptive = 1$ if the most meaningful child cluster (in any parent-child cluster pair) is dense with respect to the shape of it's parent cluster. It is not obvious as to whether this will be the case before using \clustar\ on a particular data set. Moreover, it is not obvious after the clustering has been found whether this technique has yielded the a more meaningful clustering overall. Without a recognisable way of appropriately using the $adaptive = 2$ setting\footnote{Other than perhaps when using the $({\bf v}, {\bf m})$ and $({\bf x}, {\bf v})$ feature spaces since it is evident from Fig. \ref{fig:finalmutualinfo} that the clustering power of \clustar\ is either similar or increased when using the $adaptive = 2$ setting compared to the $adaptive = 1$ on these feature spaces.}, we can only expect to see a {\it different} clustering by using it over the $adaptive = 1$ setting.

Given that using \clustar\ with the locally adaptive metric is more computationally demanding and requires a longer run-time (by a factor of $\sim2$) with varied clustering results, it is left to the user's discretion as to whether or not they wish to user it and by default the $adaptive$ parameter will be set to $1$ ensuring a globally adaptive metric. This aspect of the \clustar\ algorithm can be certainly be improved and we will endeavour to include a more appropriately determined locally adaptive metric in a future work. It is also possible to use \clustar\ on the $adaptive = 0$ setting in conjunction with other external manifold learning codes such as \code{UMAP} \citep{UMAP} -- however clustering over these results can be unpredictable and may also require a different $\rho_\mathrm{threshold}$ value than the optimal one we report in this paper.

\section{Conclusions} \label{sec:conclusion}
We have presented the \clustar\ algorithm and have shown it to be well-suited to the hierarchical classification of stellar satellite groups within galaxies. Compared to its predecessor \hoptics, \clustar\ is not only capable of producing a similarly robust clustering output but it does so in an exceptionally efficient manner -- outperforming the \hoptics\ run-times by $3$ orders of magnitude at a minimum. Importantly, \clustar\ is readily applicable to any point-based astrophysical data set that is defined over an arbitrary number of features. In optimising its clustering performance, we have removed the need for user defined parameter values such that the most optimal parameter settings will be chosen automatically based on the input data (unless the user specifies otherwise).

In this paper we also investigate the capacity of a computationally cheap design for producing a locally adaptive metric to improve \clustar's clustering power. The design iteratively re-applies \clustar's substructure searching component to each consecutive level of the hierarchy in a top-down fashion in order to adaptively modify the underlying distance metric and ensure that each level of clusters found are dense with respect to their parent cluster in the hierarchy. We find that while this approach does produce better results in a small number of clustering scenarios (predominantly on data defined with a combination of kinematic and chemical abundance features), it is mostly inconsistent with regards to the resultant clustering power and tends to simply provide a different clustering when compared to the globally adaptive metric scheme. Since the locally adaptive metric scheme requires more run-time than the globally adaptive metric scheme, we leave using it as a situational choice for the user to make.

In a future work we will develop a more clustering-appropriate locally adaptive metric scheme for \clustar\ to make use of. We will also establish a method of producing fuzzy clusterings from fuzzy data such that the point-based uncertainties of the data may be propagated into cluster-based uncertainties. As of this paper however, \clustar\ is ideally suited to large astrophysical data sets both synthetic and observational, and hence our further research will also include the application of \clustar\ to observational data sets of the MW.

\section*{Acknowledgements}
WHO gratefully acknowledges financial support through the Paulette Isabel Jones PhD Completion Scholarship at the University of Sydney. GFL received no funding to support this research.


\section*{Data Availability}
The data underlying this article may be made available on reasonable request to the corresponding author.


\bibliographystyle{mnras}
\bibliography{references}



\appendix


\bsp	
\label{lastpage}
\end{document}